\documentclass[preprint,showpacs,preprintnumbers,amsmath,amssymb,superscriptaddress,nofootinbib]{revtex4}
\usepackage{graphicx,color}
\usepackage{amsmath,amssymb}
\usepackage{url}
\usepackage{epstopdf}
\newcommand{\hs}{\hspace*{0.5cm}}

\newcommand{\be}{\begin{equation}}
\newcommand{\ee}{\end{equation}}
\newcommand{\bea}{\begin{eqnarray}}
\newcommand{\eea}{\end{eqnarray}}
\newcommand{\nn}{\nonumber}
\newcommand{\crn}{\nonumber \\}

\newcommand{\al}{\alpha}

\newcommand{\fr}{\frac}

\newcommand{\bc}{\begin{center}}
\newcommand{\ec}{\end{center}}

\newcommand {\ba}{\begin{array}}
\newcommand {\ea}{\end{array}}
\newcommand{\ben}{\begin{enumerate}}
\newcommand{\een}{\end{enumerate}}

\usepackage{bm}
\usepackage{dcolumn}

\begin{document}

\title{ Investigation of CP-even Higgs bosons decays $H \rightarrow \mu \tau$ within constraints of $l_a \rightarrow l_b \gamma$ in a 3-3-1 model with inverse seesaw neutrinos}

\author{H.V. Quyet}\email{hoangvanquyet@hpu2.edu.vn}
\affiliation{Department of Physics, Hanoi Pedagogical University 2, Phuc Yen,  Vinh Phuc 15000, Vietnam}

\author{T.T. Hieu}\email{trantrunghieu@hpu2.edu.vn}

\affiliation{Department of Physics, Hanoi Pedagogical University 2, Phuc Yen,  Vinh Phuc 15000, Vietnam}

\affiliation{Graduate Univesity of Sience and Technology, Vietnam Academy of Sience and Technology, Hanoi 10000, Vietnam}

\author{N.T. Tham}\email{nguyenthitham@hpu2.edu.vn}
\affiliation{Department of Physics, Hanoi Pedagogical University 2, Phuc Yen,  Vinh Phuc 15000, Vietnam}

\author{N.T.T. Hang}\email{hangntt@daihocpccc.edu.vn}

\affiliation{The University of fire prevention and fighting, 243 Khuat Duy Tien, Nhan Chinh, Thanh Xuan, Hanoi 10000, Vietnam}

\author{H. T. Hung\footnote{Corresponding author}}\email{hathanhhung@hpu2.edu.vn}

\affiliation{Department of Physics, Hanoi Pedagogical University 2, Phuc Yen,  Vinh Phuc 15000, Vietnam}
\affiliation{Bogoliubov Laboratory of Theoretical Physics, Joint Institute for Nuclear Research, Dubna, 141980 Russia}
\begin{abstract}
 In a 3-3-1 model with inverse seesaw neutrinos, we use a simple form of Higgs potential to give four CP-even Higgs bosons ($H \equiv h^0_1,h^0_2,h^0_3,h^0_4$). We investigate  $H \rightarrow \mu \tau$ decays in the parameter space regions satisfying the experimental limits of $l_a \rightarrow l_b \gamma$ with running parameters being the mass of the charged Higgs boson ($m_{H_1^\pm}$) and the mixing matrix of the heavy neutrinos ($M_R$). We show that there exist regions of parameter space where all partial widths $\Gamma (H \rightarrow \mu \tau)$ are less than the current experimental limit ($4.1 \times 10^{-6} GeV$). Analyzing the contributing components to $\Gamma (H \rightarrow \mu \tau)$, we also compare the mass of the SM-like Higgs boson with the corresponding ones of the other CP-even Higgs bosons in this model.
\end{abstract} 
\maketitle
 \section{\label{intro} Introduction}
 \allowdisplaybreaks
 Increasingly precise experimental data continuously confirm the non-zero mass and oscillations of neutrinos Ref.\cite{Zyla:2020zbs, Tanabashi:2018oca, Patrignani:2016xqp}. This also supports the hypothesis of the existence of  lepton-flavor-violation (LFV) processes  in the lepton part. There are two processes of great interest today, lepton-flavor-violating decays of
 charged leptons (cLFV) and lepton-flavor-violating decays of the standard model- like Higgs boson (LFVHDs). From the experimental side, branching ratio ($Br$) of cLFV have upper bound as given in Refs.\cite{Patrignani:2016xqp,ATLAS:2019erb}. 
 \bea
 Br(\mu \rightarrow e\gamma)<4.2\times 10^{-13},\,
 Br(\tau \rightarrow e\gamma)<3.3\times 10^{-8},\,
 Br(\tau \rightarrow \mu\gamma)<4.4\times 10^{-8},\label{lalb-limmit}
 \eea
 and the limits of LFVHDs are given in Refs.\cite{CMS:2018ipm,ATLAS:2019pmk,ATLAS:2019old}. 
 \bea
 Br(h \rightarrow \mu\tau)\leq \mathcal{O}(10^{-3}),\,
 Br(h \rightarrow \tau e)\leq \mathcal{O}(10^{-3}),\,
 Br(h \rightarrow \mu e)< 6.1 \times 10^{-5}. \label{hmt-limmit}
 \eea
  On the theoretical side, the LFV processes are getting more attention in the models  beyond the standard model (BSM). Some published results show that the  parameter space regions predicted from BSM for the large  signal of LFVHDs is limited directly from the experimental data of cLFV \cite{Herrero-Garcia:2016uab,Blankenburg:2012ex}. Moreover, $Br(h_1^0\rightarrow \mu\tau)$ can get values of $\mathcal{O}(10 ^ {-4})$ in supersymmetric and non-supersymmetric models \cite{Zhang:2015csm,Herrero-Garcia:2017xdu}. But,the $Br(h_1^0\rightarrow \mu\tau)$ may only get values of $\mathcal{O}(10 ^ {-9})$ \cite{Gomez:2017dhl} without the contribution of heavy neutrinos. With the addition of neutral heavy neutrinos, the models can solve two goals both creating large lepton flavor violating sources and explain the masses and mixing of neutrinos through the inverse seesaw (ISS) mechanism \cite{CarcamoHernandez:2019pmy, Catano:2012kw, Hernandez:2014lpa, Dias:2012xp}. 
 
 Recently, LFV processes are also extensively studied in the 3-3-1 models. There are two main directions to generate LFV sources, the first is to study the mass generation of neutrinos by the effective operator \cite{Mizukoshi:2010ky, Dias:2005yh,Hung:2022vqx, Hung:2022kmv}, and the second is to use the seesaw mechanism to generate the masses of neutrinos (331ISS) \cite{Boucenna:2015zwa,Hernandez:2013hea,Nguyen:2018rlb, Hung:2021fzb}. Both of these directions give good results on LFVHD, even indicating that the branching ratio can reach to $Br(h_1^0\rightarrow \mu\tau)$ is $\mathcal{O}(10 ^ {-5})$ \cite{Hue:2015fbb, Thuc:2016qva}.
 However, 331ISS model still has some questions to be solved, such as: in the parameter space regions satisfying the experimental limits of $Br(l_a \rightarrow l_b\gamma)$, is $Br(h \rightarrow l_al_b)$ excluded? What values can $\Gamma (H \rightarrow \mu \tau)$ take? Is it possible to predict the masses of the CP-even Higgs bosons? In  this work, we will solve those problems.
 
 The paper is organized as follows. In the next section, we review the model and give masses spectrum of gauge and Higgs bosons. We then show the masses spectrum of the neutrinos through the inverse seesaw mechanism in Section \ref{ISSme}. We calculate the Feynman rules and analytic formulas for cLFV and LFVHDs  in Section \ref{analytic}. Numerical results are discussed in Section.\ref{Numerical}. Conclusions are in Section \ref{conclusion}. Finally, we provide Appendix \ref{appen_loops},\ref{appen_loops1},\ref{appen_loops2},\ref{appen_loops3} to calculate the amplitude of $H \rightarrow \mu \tau$ decays.
 \section{\label{model} Brief the model}
 \subsection{Particle content}
 We are interested a model 3-3-1, which is based on the $SU(3)_L\otimes SU(3)_C\otimes U(1)_X$ gauge group with $\beta=-\frac{1}{\sqrt{3}}$. So, electric charged operator is defined $Q=T_3+\beta T_8+X$, where $T_{3,8}$ are diagonal $SU(3)_L$ generators. In this model, right-handed neutrinos must accommodate at the bottom of the lepton triplets, which are the heavy neutrinos, and to give mass to the active neutrinos rule the inverse seesaw mechanism, we need to add gauge singlets $F_a$. In this way, we do not need the right-handed singlets of active and heavy neutrinos.
The fermions in the model are arranged as Eq.(\ref{f_content}). These options are both simple and guarantee chiral anomalies cancelling as mentioned in Ref.\cite{Diaz:2004fs}. 
\begin{align} \label{f_content}
L'_{aL}& =%
\begin{pmatrix}
\nu' _{a} \\ 
l'_{a} \\ 
\left( N' _{a}\right) ^{c} \\ 
\end{pmatrix}%
_{L}:(1,3,-1/3), & & 
l'_{aR}:(1,1,-1) %
,  \notag \\
Q'_{\al L}& =%
\begin{pmatrix}
d'_{\al} \\ 
-u'_{\al} \\ 
D'_{\al} \\ 
\end{pmatrix}%
_{L}:(3,3^{\ast },0), & & 
\begin{cases}
d'_{\al R}:(3,1,-1/3) \\ 
u'_{\al R}:(3,1,2/3) \\ 
D'_{\al R}:(3,1,-1/3) \\ 
\end{cases}%
,  \notag \\
Q_{3L}^{\prime}& =%
\begin{pmatrix}
u'_{3} \\ 
d'_{3} \\ 
U'_3 \\ 
\end{pmatrix}%
_{L}:(3,3,1/3), & & 
\begin{cases}
u'_{3R}:(3,1,2/3) \\ 
d'_{3R}:(3,1,-1/3) \\ 
U'_{3R}:(3,1,2/3) \\ 
\end{cases},%
\end{align}
where $U'_{3L}$ and $D'_{\al L}$ for $\al=1,2$ are up- and down-type quark components in the flavor basis, $N_{aL}^{\prime c} \cong N'_{aR}$ are right-handed neutrinos added in the bottom of lepton triplets, the quantum numbers in parentheses are the gauge charged of the groups $(SU(3)_C, SU(3)_L, U(1)_X)$, respectively. Commas are used to distinguish the initial state from the physical states.\\
Three scalar triplets $\eta, \rho, \chi $ are given to provide the  masses of particles through the expectation vacuum values (VEVs). They are assigned to the following $(SU(3)C; SU(3)L; U(1)X) $  representations. 
\begin{eqnarray}
\eta =%
\begin{pmatrix}
\eta_1^0   \\ 
\eta _{2}^{-} \\ 
\eta_3^0%
\end{pmatrix}%
:(1,3,-1/3);
\hs
\rho =%
\begin{pmatrix}
\rho_{1}^{+} \\ 
\rho_2^0  \\ 
\rho _{3}^{+} \\ 
\end{pmatrix}%
:(1,3,2/3);  \hs
\chi =%
\begin{pmatrix}
\chi_1^0 \\ 
\chi_{2}^{-} \\ 
\chi_3^0  \\ 
\end{pmatrix}%
:(1,3,-1/3), 
\label{scalar_spectrum1}
\end{eqnarray}
and their VEVs have form, 
\bea 
\eta_1^0 &=&\frac{1}{\sqrt{2}}(v_1+R_{1 }+ iI_{1 }); \hs \eta_3^0= \frac{1}{\sqrt{2}}(R_{1 }^\prime+ iI_{1 }^\prime) \crn
\rho_2^0 &=&\frac{1}{\sqrt{2}}(v_2+R_{2 }+ iI_{2}); \hs \chi_1^0 = \frac{1}{\sqrt{2}}(R_{3 }^\prime+ iI_{3 }^\prime) ; \hs
\chi_3^0 = \frac{1}{\sqrt{2}}(v_3 +R_{3 }+ iI_{3 })  \label{scalar_spectrum2},
\eea
where $\chi$ takes care of breaking the $SU(3)_{L}\otimes U(1)_{X}$ symmetry group to give masses to heavy particles, which are essentially outside of standard model, $\rho$ gives mass to charged leptons and a down-type and two up-type quarks, $\eta$ gives mass to the remaining quarks Refs.~\cite{Chang:2006aa,Tully:2000kk}. Active neutrinos are mass generated by the inverse seesaw mechanism \cite{Nguyen:2018rlb, Hung:2021fzb}. Therefore, the electroweak symmetry breaking (EWSB) mechanism is ruled follows,
\begin{equation*}
{SU(3)_{L}\otimes U(1)_{X}\xrightarrow{\langle \chi \rangle}}{%
	SU(2)_{L}\otimes U(1)_{Y}}{\xrightarrow{\langle \eta \rangle,\langle \rho
		\rangle}}{U(1)_{Q}},
\end{equation*}
and VEVs satisfy the hierarchy ${v_3 \gg v_1,v_2}$  as done
in Refs. \cite{Dong:2008sw,Dong:2010gk}.   

 Because $\chi$ and $\eta$ have the same quantum numbers, they perform the same formal role in the mass mixing matrix ($M_R$). We can use the ($\chi +\kappa \eta$) combination in the corresponding Lagrangian term. The Lagrangian relates to leptons and heavy neutrinos are given follows: 
\bea 
-\mathcal{L}_{LF} =h^l_{ab}\overline{L'_{aL}}\rho l'_{bR}-
h^{\nu}_{ab} \epsilon^{ijk} \overline{(L'_{aL})_i}(L'_{bL})^c_j\rho^*_k+h^N_{ab}\overline{L'_{aL}}\left( \chi +\kappa \eta \right)  F'_{bR}+\frac{1}{2} (\mu_{F})_{ab}\overline{(F'_{aR})^c}F'_{bR}
+\mathrm{H.c.}\label{Yul}
\eea 
The first term in Eq.(\ref{Yul})  generate masses for original charged leptons, the second one give mass mixing matrix ($M_D$) and the next term describes mixing between $N'_a$ and $F'_a$, and the last term generates masses for Majorana neutrinos $F'_a$. Based on the hierarchy ${v_3 \gg v_1,v_2}$ and to reduce the number of degrees of freedom in the model, we can ignore the contribution of $\eta$ to the matrix ($M_R$). That means $\kappa =0$. The masses of neutrinos obtained through the inverse seesaw mechanism will be shown later.
\subsection{Higgs and Gauge bosons} 
We use the simple form of the Higgs potential as discussed in Refs. \cite{Hue:2015mna, Hue:2015fbb} to gain the mass and state of  SM-like Higgs boson which will be determined exactly at the tree level. That form is, 
\bea
\mathcal{V}_H &=& \mu_1^2 \left( \rho^\dagger \rho + \eta^\dagger \eta \right) + \mu_2^2 \chi^\dagger \chi + \lambda_1 \left( \rho^\dagger \rho + \eta^\dagger \eta \right)^2 + \lambda_2 \left( \chi^\dagger \chi \right)^2 + \lambda_{12}  \left( \rho^\dagger \rho + \eta^\dagger \eta \right)  \left( \chi^\dagger \chi \right)
\crn
&&- \sqrt 2 fv_3 \left(\varepsilon_{ijk} \eta^i \rho^j \chi^k + \mathrm{H.c.}  \right),
\label{potential}
\eea
where $\lambda_1,\, \lambda_2,\, \lambda_{12}$ are the  Higgs self-coupling constants, $f$ is a dimensionless coefficient and assumed to be real. \\
This model has two single-charged Higgs bosons, whose states are mixed with the two Goldstone bosons of the gauge bosons $W^\pm$ and $V^\pm$ in the following form: 
\begin{eqnarray}
\left( \begin{array}{c}
\rho_1^\pm \\
\eta_2^\pm
\end{array} \right)= \dfrac{1}{\sqrt 2}\left( \begin{array}{cc}
-1 & 1\\
1 & 1
\end{array} \right) \left(\begin{array}{c}
G_W^\pm
\\ H_1^\pm
\end{array} \right), \hs \left( \begin{array}{c}
\rho_3^\pm \\
\chi_2^\pm
\end{array} \right) = \left( \begin{array}{cc}
- s_{23} & c_{23} \\
c_{23} & s_{23}
\end{array} \right) \left(\begin{array}{c}
G_V^\pm
\\ H_2^\pm
\end{array} \right),
\label{EchargedH}
\end{eqnarray}
and their masses
\bea 
&& m_{H_1^{\pm}}^2=2fv_3^2,\hs m_{H_2^{\pm}}^2=fv_3^2(t^2_{23} +1),\crn
\text{with}&& \hs c_{23} \equiv \cos\beta_{23},\hs s_{23} \equiv \sin\beta_{23}, \hs t_{23} \equiv \tan{\beta_{23}}=\frac{v_2}{v_3}.
\eea 
In addition, the model contains four physical CP-even Higgs bosons $h^0_{ 1;2;3;4}$ based on components of scalar fields are constructed as Eq.(\ref{scalar_spectrum2}). Among of them, a neutral Higgs boson ($h_4^0$) mix with a Goldstone of boson $X^0$ which depends on $\beta_{13}$ angle, ($t_{13} \equiv \tan\beta_{13} =\frac{v_1}{v_3}$).
\begin{eqnarray}
 \left( \begin{array}{c}
R_1^\prime \\
R_3^\prime
\end{array} \right) = \left( \begin{array}{cc}
- s_{13} & c_{13} \\
c_{13} & s_{13}
\end{array} \right) \left(\begin{array}{c}
G_{X}
\\ h_4^0
\end{array} \right)\hs \text{and} \hs  m_{h_4^{0}}^2=fv_3^2(t^2_{13} +1).
\label{mixh04}
\end{eqnarray}

In the original basis ($R_1,R_2.R_3$), the mixing matrix of the squared mass of three neutral Higgs bosons ($h^0_1,h^0_2,h^0_3$) is:
\bea
\mathcal{M}_h^2=\left(
\begin{array}{ccc}
	2 \lambda _1 v_1^2+f v_3^2 & 2 v_1^2 \lambda _1-f v_3^2 & v_1 v_3 \left(\lambda _{12}-f\right) 
	\\
	2 v_1^2 \lambda _1-f v_3^2 & 2 \lambda _1 v_1^2+f v_3^2 & v_1 v_3 \left(\lambda _{12}-f\right)
	\\
	v_1 v_3 \left(\lambda _{12}-f\right) & v_1 v_3 \left(\lambda _{12}-f\right) & f v_1^2+2 v_3^2 \lambda
	_2 \\
\end{array}
\right)
\eea
Their respective masses obtained are:
\bea
m_{h_1^0}^2 &&= \frac{1}{2} \left(f v_1^2+4 \lambda _1 v_1^2+2 \lambda _2 v_3^2 \right. \crn
&&\left.-\sqrt{8 v_1^2 \left(f^2 v_3^2-2 f \left(\lambda _1 v_1^2+\lambda _{12}
	v_3^2\right)+\left(\lambda _{12}^2-4 \lambda _1 \lambda _2\right) v_3^2\right)+\left(f v_1^2+4 \lambda _1 v_1^2+2 \lambda _2
	v_3^2\right){}^2}\right)\crn
m_{h_2^0}^2&& = \frac{1}{2} \left(f v_1^2+4 \lambda _1 v_1^2+2 \lambda _2 v_3^2\right.\crn
&&\left.+\sqrt{8 v_1^2 \left(f^2 v_3^2-2 f \left(\lambda _1 v_1^2+\lambda _{12}
	v_3^2\right)+\left(\lambda _{12}^2-4 \lambda _1 \lambda _2\right) v_3^2\right)+\left(f v_1^2+4 \lambda _1 v_1^2+2 \lambda _2v_3^2\right){}^2}\right)\crn
m_{h_3^0}^2&& = 2fv^2_3. \label{massesnHiggs}
\eea
 Accordingly, the relationship between the physical states and the initial state is also given,
\bea
\left(\begin{array}{c} R_1\\ R_2\\ R_3 \end{array}\right) =
\left(
\begin{array}{ccc}
	-\fr{c_\alpha}{\sqrt{2}} & \fr{s_\alpha}{\sqrt{2}} & -\fr{1}{\sqrt{2}} \\
	-\fr{c_\alpha}{\sqrt{2}} & \fr{s_\alpha}{\sqrt{2}} & \fr{1}{\sqrt{2}} \\
	s_\alpha & c_\alpha & 0 \\
\end{array}
\right)\left(\begin{array}{c} h^0_1\\ h^0_2\\ h^0_3 \end{array}\right),
\label{mixing_CP_even_Higgs}
\eea
where $s_\alpha = \sin\alpha$ and $c_\alpha = \cos\alpha$, and they are defined by
\bea
s_\alpha &=& \fr{(4 \lambda_1-m^2_{h^0_1}/v^2)t_{13} }{A},\,
c_\alpha= \fr{\sqrt{2}\left(\lambda_{12}-f\right)}{A},\crn  A&=&\sqrt{ \left(4 \lambda_1 -m^2_{h^0_1}/v^2\right)^2 t_{13}^2+2\left(\lambda_{12}-f\right)^2}.
\label{angle_beta}
\eea
The neutral CP-even Higgs boson $h^0_1$ is the lightest and identified with SM-like Higgs boson. we can give its mass in a more compact form \cite{Hung:2021fzb, Nguyen:2018rlb}.

\be m^2_{h^0_1}=\frac{v_3^2}{2}\left[4\lambda_1 t_{13}^2 + 2\lambda_2 +ft_{13}^2-\sqrt{8 t^2_{13}\left(f-\lambda_{12}\right)^2+\left(2\lambda_2 +ft_{13}^2-4 \lambda_1 t^2_{13}\right)^2} \right]. \ee

Gauge bosons in this model get masses through the covariant kinetic term of the Higgs bosons,
\bea \mathcal{L}^{H}=\sum_{H=\eta,\rho,\chi} \left(D_{\mu}H\right)^{\dagger}\left(D_{\mu}H\right).\nn\eea

The model comprises two pairs of singly charged gauge bosons, denoted as  $W^{\pm}$  and $V^{\pm}$, defined as
\bea W^{\pm}_{\mu}&=&\frac{W^1_{\mu}\mp i W^2_{\mu}}{\sqrt{2}},\hs m_W^2=\frac{g^2}{4}\left(v_1^2+v_2^2\right),\crn
V^{\pm}_{\mu}&=&\frac{W^6_{\mu}\pm i W^7_{\mu}}{\sqrt{2}},\hs m_V^2=\frac{g^2}{4}\left(v_2^2+v_3^2\right). \label{singlyG}\eea
The bosons $W^{\pm}$ as the first line in Eq.(\ref{singlyG}) are identified with the SM ones, leading to $v_1^2+v_2^2\equiv v^2=(246\, \mathrm{GeV})^2$.  In the remainder of the text, we will consider in detail the simple case $v_1=v_2=v/\sqrt{2}=\sqrt{2}m_W/g$ given in Refs. \cite{Hue:2015mna, Hue:2015fbb, Thuc:2016qva}. Under these imposing conditions, we get $h^0_1$ mixed with the three original states.
\section{\label{ISSme} Neutrinos masses and ISS mechanism} 
We are now interested in generating the masses for neutrinos through the inverse seesaw mechanism. The mixing matrix of active neutrinos with heavy neutrinos is derived from the  second term in Eq. (\ref{Yul}) which is expanded as following:
  \begin{align}
 	h^{\nu}_{ab} \epsilon^{ijk} \overline{(L'_{aL})_i}(L'_{bL})^c_j\rho^*_k
 	= 2 h^{\nu}_{ab} \left[-\overline{l'_{aL}} (\nu'_{bL})^c\rho^-_3+ \overline{l'_{aL}} (N'_{bL})^{c}\rho^-_1-\overline{\nu'_{aL}} (N'_{bL})^c\rho_2^{0*} \right]. \label{lastterm}
 \end{align}
We pay attention to the last term of Eq.(\ref{lastterm}) which is the term that creates the $m_D$ matrix. Using  antisymmetric properties of $h^\nu_{ab}$  matrix and equality $\overline{N_{aL}} (\nu_{bL})^c=\overline{\nu_{bL}} (N_{aL})^c$, we can contribute a Dirac neutrino mass term $-\mathcal{L}_{\rm{mass}}^{\nu}=\overline{\nu_L}\,m_D\, N_R+\rm{H.c.}$, in the basic, $\nu_L\equiv(\nu_{1L},\nu_{2L},\nu_{3L})^T$, $N_R\equiv( (N_{1L})^c, (N_{2L})^c,(N_{3L})^c)^T$ where $(m_D)_{ab}\equiv \sqrt{2}\, h^{\nu}_{ab}v_2$ with $a,b=\rm{1,2,3.}$\\
The third term in Eq.(\ref{Yul}) generates mass for heavy neutrinos, this consequence comes from the large value of Yukawa coupling $Y_{ab}$. To describe mixing $N'_a$ and $F'_a$, $(M_R)_{ab}=h^N_{ab}\frac{v_3}{\sqrt{2}}$ is introduced. To avoid lepton number violation as mentioned in Refs.\cite{Hung:2021fzb, Nguyen:2018rlb}, hence $\mu_F$ can be assumed to
be small, in the scale of ISS models.\\
 In the basis $n'_L=(\nu_L,N_{L},(F_{R})^c)^T$ and $(n'_L)^c=((\nu_L)^c,(N_{L})^c,F_{R})^T$, Eq.(\ref{Yul}) derives mass matrix following. 
 \bea -{L}^{\nu}_{\mathrm{mass}}=\frac{1}{2}\overline{n'_L}M^{\nu}(n'_L)^c +\mathrm{H.c.}, \,\mathrm{ where }\quad M^{\nu}=\begin{pmatrix}
 	0	& m_D &0 \\
 	m^T_D	&0  & M_R^T \\
 	0& M_R& \mu_F
 \end{pmatrix},  \label{Lnu1}\eea
In the normal seesaw form, the matrix $M^\nu$ can be written in term of:
 \be  M^{\nu}=\begin{pmatrix}
 	0& M_D \\
 	M_D^T& M_N
 \end{pmatrix}, \; \mathrm{where} \, M_D \equiv(m_D,\, 0),\;  \mathrm{and} \; M_N=\begin{pmatrix}
 	0& M_R^T \\
 	M_R& \mu_F
 \end{pmatrix}. \label{Mnuss}\ee
The matrix $M^\nu$ can be diagonalized by a $9\times9$ unitary matrix $U^\nu$ to obtain mass eigenvalues and physics states of neutrinos,
\bea U^{\nu T}M^{\nu}U^{\nu}=\hat{M}^{\nu}=\mathrm{diag}(m_{n_1},m_{n_2},..., m_{n_{9}})=\mathrm{diag}(\hat{m}_{\nu}, \hat{M}_N), \label{diaMnu} \eea
where $m_{n_k}$($k=1,2,...,9$) are  masses of the nine physical neutrino states $n_{kL}$. They consist of three active neutrinos $n_{aL}$ ($a=1,2,3$) corresponding to mass  submatrix    $\hat{m}_{\nu}=\mathrm{diag}(m_{n_1},\;m_{n_2},\;m_{n_3})$, and the six extra neutrinos $n_{jL}$ ($j=4,5,..,9$)  with  $\hat{M}_N=\mathrm{diag}(m_{n_4},\;m_{n_5},...,\;m_{n_{9}})$. \\
The relations between the flavor and mass eigenstates of neutrinos are:
\bea n'_L=U^{\nu*} n_L, \hs \mathrm{and} \; (n'_L)^c=U^{\nu}  (n_L)^c, \label{Nutrans}
\eea
\be P_Ln'_k=n'_{kL} =U^{\nu*}_{kj}n_{jL},\; \mathrm{and}\; P_Rn'_k=n'_{kR} =U^{\nu}_{kj}n_{jR}, \hs k,j=1,2,...,9. \label{Nutrans2}\ee
Based on the ISS mechanism, $U^{\nu}$ can be written in the form   \cite{Ibarra:2010xw},
\be U^{\nu}= \Omega \left(
\begin{array}{cc}
	U^0_{\rm{PMNS}} & \mathbf{O} \\
	\mathbf{O} & V \\
\end{array}
\right), \hs
\label{Unuform}\ee
\be \Omega=\exp\left(
\begin{array}{cc}
	\mathbf{O} & R \\
	-R^\dagger & \mathbf{O} \\
\end{array}
\right)=
\left(
\begin{array}{cc}
	1-\frac{1}{2}RR^{\dagger} & R \\
	-R^\dagger &  1-\frac{1}{2}R^{\dagger} R\\
\end{array}
\right)+ \mathcal{O}(R^3).
\label{Ommatrix}\ee
where the matrix $U^0_{\mathrm{PMNS}}$ in Eq.(\ref{Unuform}) is the Pontecorvo-Maki-Nakagawa-Sakata (PMNS) matrix, and following seesaw relations \cite{Ibarra:2010xw} are :
\begin{align}
	R^* &\simeq \left(-m_DM^{-1}, \quad  m_D(M_R^\dagger)^{-1}\right), \quad  M\equiv M_R\mu_F^{-1}M_R, \label{Rs}\\
	m_DM^{-1} m^T_D&\simeq m_{\nu}\equiv U^*_{\mathrm{PMNS}}\hat{m}_{\nu}U^{\dagger}_{\mathrm{PMNS}},  \label{mnu}\\
	V^* \hat{M}_N V^{\dagger}& \simeq  M_N+ \frac{1}{2}R^TR^* M_N+ \frac{1}{2} M_NR^{\dagger} R.\;
	 \label{masafla}
\end{align}
The standard form of the lepton mixing matrix $U_{\rm{PMNS}}$ is the function of three angles $\theta_{ij}$, one Dirac phase $\delta$ and two Majorana phases $\gamma_{1}$,and $\gamma_{2}$, namely \cite{Tanabashi:2018oca}
\begin{align}
	U_{\mathrm{PMNS}}
	&= \begin{pmatrix}
		1	& 0 &0  \\
		0	&c'_{23}  &s'_{23}  \\
		0&  	-s'_{23}& c'_{23}
	\end{pmatrix}\,\begin{pmatrix}
		c'_{13}	& 0 &s'_{13}e^{-i\delta}  \\
		0	&1  &0  \\
		-s'_{13}e^{i\delta}&  0& c'_{13}
	\end{pmatrix}\,\begin{pmatrix}
		c'_{12}	& s'_{12} &0  \\
		-s'_{12}	&c'_{12}  &0  \\
		0& 0 	&1
	\end{pmatrix} \mathrm{diag}\left(1, e^{i\gamma_{1}},\,e^{i\gamma_{2}}\right) 
	\crn&=U^0_{\mathrm{PMNS}} \;\mathrm{diag}\left(1, e^{i\gamma_{1}},\,e^{i\gamma_{2}}\right),\;\label{umns}
\end{align}
in which $s'_{ij}\equiv\sin\theta_{ij}$, $c'_{ij}\equiv\cos\theta_{ij}=\sqrt{1-s^2_{ij}}$, $i,j=1,2,3$ ($i<j$), $0\le \theta_{ij}<90\; [\mathrm{Deg.}]$ and $0<\delta\le 720\;[\mathrm{Deg.}].$\\
The  Dirac phase ($\delta$) and Majorana phases ($\gamma_1,\gamma_2$) are fixed as $\delta=\pi,\gamma_1=\gamma_2=0$. In the normal hierarchy scheme, the respective best-fit values of the neutrino oscillation parameters which satisfied the confidence level of $3\sigma$ are given as  \cite{Zyla:2020zbs}
%
\bea 
s'^2_{12}&=&0.32,\; s'^2_{23}=0.551,\; s'^2_{13}=0.0216,\crn
\Delta m^2_{21}&=& 7.55\times 10^{-5}\;\mathrm{ eV^2},\hs  \Delta m^2_{32}= -2.50\times 10^{-3}\; \mathrm{eV^2},\label{nuosc}
\eea
where $ \Delta m^2_{21}=m^2_{n_2}-m^2_{n_1}$ and $\Delta m^2_{32}=m^2_{n_3}-m^2_{n_2}$. \\
In this paper, we work on the framework of the 331RHN model with added flavor singlets $F_a$, the Dirac mass matrix of neutrinos $m_D$ must be antisymmetric. From results in Ref.\cite{Boucenna:2015zwa}, with the aim of finding regions of parameter space with large LFVHDs, the matrix $m_D$ can be chosen to satisfy both the inverse and normal hierarchy cases of active neutrino masses, as the following expressed: 
\be m_D\equiv \varrho \begin{pmatrix}
	0&1  &x_{13}  \\
	-1& 0 &x_{23}  \\
	-x_{13}& -x_{23} &0
\end{pmatrix}, \label{mD1}\ee
as a result, $m_D$ depends on three parameters only, $x_{13},x_{23}$, and $\varrho=\sqrt{2}vh^\nu_{23}$; $\varrho$ is assumed to be positive and real.\\
In general, the matrix $m_{\nu}$ in Eq.(\ref{mnu}) is symmetric, $(m_{\nu})_{ab}=(m_{\nu})_{ba}$ ($a,b=\rm{1,2,3}$), their components are given by $(m_\nu)_{ab}=(m_D)_{ai}(M^{-1})_{ij}(m_D^T)_{jb},\quad (a\ne i, j\ne b)$. Calculating in detail
\bea 
(m_{\nu})_{ab}-(m_{\nu})_{ba}&\sim x_{13}\left[M^{-1}_{13}-M^{-1}_{31}\right]+ x_{23}\left[M^{-1}_{23}-M^{-1}_{32}\right]+(M^{-1}_{12}-M^{-1}_{21}),\quad \text{for}\quad  a\ne b \crn
(m_\nu)_{11}&=M^{-1}_{22}+x_{13}(M^{-1}_{23}+M^{-1}_{32})+x^2_{13}M^{-1}_{33},\crn
(m_\nu)_{22}&=-M^{-1}_{11}-x_{23}(M^{-1}_{13}+M^{-1}_{31})+x^2_{23}M^{-1}_{33},\crn
(m_\nu)_{33}&=x^2_{13}M^{-1}_{11}+x_{13}x_{23}(M^{-1}_{12}+M^{-1}_{21})+x^2_{23}M^{-1}_{22}.
\label{mnu_ij}
\eea   
From Eq.(\ref{mnu_ij}), we have two solutions $x_{13}$, $x_{23}$ and one equation, which expresses the relation between elements of matrix $m_\nu$: 
\bea x_{13}&=\frac{(m_{\nu})_{23}\left[ (m_{\nu})_{13}^2-(m_{\nu})_{11}(m_{\nu})_{33}\right] +(m_{\nu})_{13}\sqrt{\left[ (m_{\nu})_{13}^2-(m_{\nu})_{11}(m_{\nu})_{33}\right] \left[ (m_{\nu})_{23}^2-(m_{\nu})_{22}(m_{\nu})_{33}\right] }}{(m_{\nu})_{13}^2(m_{\nu})_{22}-(m_{\nu})_{11}(m_{\nu})_{23}^2},\crn
x_{23}&=\frac{(m_{\nu})_{13}\left[ (m_{\nu})_{23}^2-(m_{\nu})_{22}(m_{\nu})_{33}\right] +(m_{\nu})_{23}\sqrt{\left[ (m_{\nu})_{13}^2-(m_{\nu})_{11}(m_{\nu})_{33}\right] \left[ (m_{\nu})_{23}^2-(m_{\nu})_{22}(m_{\nu})_{33}\right] }}{(m_{\nu})_{13}^2(m_{\nu})_{22}-(m_{\nu})_{11}(m_{\nu})_{23}^2},\crn
(m_{\nu})_{11}&(m_{\nu})^2_{23}+(m_{\nu})_{22}(m_{\nu})^2_{13}+ (m_{\nu})_{33}(m_{\nu})^2_{12}=
(m_{\nu})_{11}(m_{\nu})_{22}(m_{\nu})_{33}+ 2(m_{\nu})_{12}(m_{\nu})_{13}(m_{\nu})_{23}.\crn
\label{nxij}\eea
From experimental data on neutrino oscillation as expressed in Eq.(\ref{nuosc}), the matrix $m_D$ is parameterized and then only depends on the parameter $\varrho$.
\bea  m_D\simeq \varrho \times \begin{pmatrix}
	0	& 1 & 0.7248 \\
	-1	& 0 & 1.8338 \\
	-0.7248& -1.8338 &0
\end{pmatrix}, \label{nmD}\eea
the Dirac matrix $m_D$ in Eq.(\ref{nmD}) is now well-suited for investigating LFVHDs.
\section{\label{analytic} Couplings and analytic formulas} 
We will now give couplings and analytic formulas in terms of $M^\nu$ and the physical states of the particles. With this aim, all vertices are presented in term of physical masses and mixing parameters. From Eq.(\ref{diaMnu}), we have:
\bea
M^{\nu}_{ab}&=&\left(U^{\nu*}\hat{M}^{\nu}U^{\nu\dagger}\right)_{ab}=0 \rightarrow U^{\nu*}_{ak}U^{\nu*}_{bk}m_{n_k}=0,\label{conMnu}
\eea
where $a,b=1,2,3$ and $m_{n_k}$ is mass of neutrino $n_k$, with $k$ run taken over $1,2,...,9$. The result in Eq.(\ref{conMnu}) leads to represent Yukawa couplings in term of $M^\nu$ and physical neutrino masses.
\bea
\sqrt{2}v_2\,h^{\nu}_{ab} &=& (m_D)_{ab}= (M^{\nu})_{a(b+3)}=(U^{\nu*}\hat{M}^{\nu}U^{\nu\dagger})_{a(b+3)}=U^{\nu*}_{ak}U^{\nu*}_{(b+3)k}m_{n_k},\crn
\frac{v_3}{\sqrt{2}} h^N_{ab}&=&(M_R)_{ab}= (M^{\nu})_{(a+3)(b+6)}=U^{\nu*}_{(a+3)k}U^{\nu*}_{(b+6)k}m_{n_k}.
\label{rel1}\eea
We then use Lagrangian Yukawa, Lagrangian kinetics of lepton (or scalar) fields and Higgs potential to give couplings relating to LFVHD. From the first term in Eq.(\ref{Yul}), we have couplings between Higgs boson and charged leptons as follow:
\bea
&& -h^l_{ab}\overline{L}_{aL}\rho l'_{bR}+{\rm h.c.}= -\frac{g  m_{l_a}}{m_W}\left( \overline{l_{aL}'}l_{aR}'\rho_2^{0}+\overline{\nu_{aL}'}l_{aR}'\rho^+_1 + \overline{N_{aL}'}l_{aR}'\rho^+_3+\rm{ h.c.}\right)\crn
&\supset&-\frac{g\,m_{l_a}}{2m_W}\left(  -c_\alpha h^0_1+s_\alpha h^0_2+h^0_3\right)\overline{l_{a}}l_{a} -\frac{g\, m_{l_a}}{\sqrt{2} m_W} \left(U^{\nu}_{ai} \overline{n_{i}}P_Rl_{a}H^+_1+ U^{\nu*}_{ai} \overline{l_{a}}P_Ln_{i}H^-_1\right)\crn
&&-\frac{g\, m_{l_a}\,c_{23}}{m_W}\left(U^{\nu}_{(a+3)i} \overline{n_{i}}P_Rl_{a}H^+_2+ U^{\nu*}_{(a+3)i} \overline{l_{a}}P_Ln_{i}H^-_2\right).
\label{eephi}\eea
 The couplings in the second term of the Lagrangian in Eq. (\ref{Yul}) are
\bea
&& h^{\nu}_{ab} \epsilon^{ijk} \overline{(L_{aL})_i}(L_{bL})^c_j\rho_k^*+ \rm{h.c.}\crn
&=&2 h^{\nu}_{ab} \left[\overline{(N_{aR}')^c} (l_{bL}')^c\rho^-_1-\overline{l_{aL}'} (\nu_{bL}')^c\rho^-_3-\overline{\nu_{aL}'} N_{bR}'\rho_2^{0*} \right]+ \rm{h.c.}\crn
&=&\frac{g\, m_{l_a}}{m_W}\left(  c_\alpha h^0_1-s_\alpha h^0_2-h^0_3\right)\left[ \sum_{c=1}^3U^{\nu}_{ci}U^{\nu*}_{cj}\overline{ n_i}\left(m_{n_i}P_L+m_{n_j}P_R\right)n_j \right]\crn
&+&\frac{g}{\sqrt{2}m_W} \left[ (m_D)_{ab}U^{\nu}_{(b+3)i} H^-_1\overline{l_{a}}P_Rn_i+\rm{h.c.}\right]\crn
&-&\frac{gc_{23}}{m_W} \left[ (m_D)_{ab}U^{\nu}_{bi} H^-_2\overline{l_{a}}P_Rn_i+\rm{h.c.}\right].
\label{psipsiphi}\eea
As mentioned above, $\chi$ and $\eta$ have the same quantum numbers so they play a similar role in generating $M_R$. However, for simplicity we choose $\kappa = 0$, then the third term of Eq.Eq. (\ref{Yul}) is expanded as :
\bea
 &-& h^N_{ab} \overline{L'_{aL}}\,\chi F'_{bR}+\rm{h.c.} = - h^N_{ab}\left( \overline{\nu'_{aL}}\chi_1^{0} + \overline{l'_{aL}}\chi^-_2 + \overline{N'_{aL}}\chi^0_3\right) F'_{bR} +\rm{h.c.}\crn
&\supset& -h^N_{ab}\left( U^{\nu}_{(a+3)i} U^{\nu}_{(b+6)j}\overline{n_{i}}P_Rn_{j}\left( s_\alpha h^0_1+c_\alpha h^0_2\right) + s_{23}U^{\nu}_{(b+6)i} \overline{l_{a}}P_Rn_{i}H^-_2 +\rm{h.c.} \right), \label{NRYe}\eea
Using the kinetic terms of the leptons, we can give couplings of the charged gauge bosons and leptons, the results are.
\bea  \mathcal{L}^{\ell\ell V}=\overline{L'_{aL}}\gamma^{\mu}D_{\mu}L'_{aL}
&\supset&\frac{g}{\sqrt{2}} \left( \overline{l'_{aL}}\gamma^\mu  \nu'_{aL}W^{-}_{\mu} + \overline{l'_{aL}}\gamma^\mu  N'_{aL} V^{-}_{\mu} \right)+\mathrm{h.c.}\crn
&=&\frac{g}{\sqrt{2}} \left[ U^{\nu*}_{ai} \overline{l_{a}}\gamma^\mu P_L  n_{i}W^{-}_{\mu} + U^{\nu}_{ai} \overline{ n_{i}}\gamma^\mu P_L l_{a}W^{+}_{\mu}\right.\crn
&+& \left.U^{\nu*}_{(a+3)i}\overline{l_{a}}\gamma^\mu  P_L n_{i} V^{-}_{\mu} + U^{\nu}_{(a+3)i}\overline{n_{i} }\gamma^\mu  P_L l_{a}V^{+}_{\mu} \right], \label{llv1}\eea
We can define symmetry coefficient $\lambda^H_{ij}=\lambda^H_{ji}$ to use for $H\overline{ n_{i}}n_i$ couplings, the results therefore obtained.
\bea \lambda^{h^0_1}_{ij}&&=-\sum_{k=1}^3\left(U^{\nu}_{ki}U^{\nu*}_{kj}m_{n_i}+U^{\nu*}_{ki}U^{\nu}_{kj}m_{n_j}\right) \crn
&&+t_\alpha t_{13}\sum_{k,q=1}^3
(M^*_R)_{cd} \left[U^{\nu*}_{(k+3)i} U^{\nu*}_{(q+6)j}+U^{\nu*}_{(k+3)j} U^{\nu*}_{(q+6)i}\right], \crn
\lambda^{h^0_2}_{ij}&&=\sum_{k=1}^3\left(U^{\nu}_{ki}U^{\nu*}_{kj}m_{n_i}+U^{\nu*}_{ki}U^{\nu}_{kj}m_{n_j}\right) \crn
&&+\frac{t_{13}}{t_\alpha } \sum_{k,q=1}^3
(M^*_R)_{cd} \left[U^{\nu*}_{(k+3)i} U^{\nu*}_{(q+6)j}+U^{\nu*}_{(k+3)j} U^{\nu*}_{(q+6)i}\right]  ,\crn
\lambda^{h^0_3}_{ij}&&= \sum_{k=1}^3\left(U^{\nu}_{ki}U^{\nu*}_{kj}m_{n_i}+U^{\nu*}_{ki}U^{\nu}_{kj}m_{n_j}\right),\crn
\lambda^{h^0_4}_{ij}&&=0.
\label{laijs}
\eea
The coefficients related to the interaction of charged Higgs and fermions as follows:
 \bea \lambda^{L,1}_{ak}&=& -\sum_{i=1}^3 (m_D^*)_{ai}U^{\nu*}_{(i+3)k} ,\quad 
 \lambda^{R,1}_{ak}=m_{l_a}U^{\nu}_{ak},
\crn
\lambda^{L,2}_{ak}&=& \sum_{i=1}^3\left[(m_D^*)_{ai}U^{\nu*}_{ik}+ t_{13}^2(M_R^*)_{ai}U^{\nu*}_{(i+6)k}\right],\quad \lambda^{R,2}_{ak}=m_{l_a}U^{\nu}_{(a+3)k}.
\eea
The couplings are listed in Tab.(\ref{numbers}). We also note some others that equal zero and is not included here, such as : $h_1^0H_1^\pm H_2^\mp$,\,$h_1^0Y^\pm W^\mp$,\,$h_1^0Y^\pm H_1^\mp$,\,$h_1^0W^\pm H_{1,2}^\mp$ ... 
  \begin{table}[ht]
	\begin{tabular}{|c|c|}
		\hline
		Vertex & Coupling \\
		\hline
		$ W_\mu^+ \overline{ n_k} l_b$,  	$   W_\mu^-\overline {l_a}n_k$ & $\frac{ig}{\sqrt{2}}U^{\nu}_{bk}\gamma^\mu P_L$, $\frac{ig}{\sqrt{2}}U^{\nu*}_{ak}\gamma^\mu P_L$\\
		\hline
		$V_\mu^+ \overline{n_k}  l_b$, 	$V_\mu^-  \overline {l_a} n_k $& $\frac{ig}{\sqrt{2}}U^{\nu}_{(b+3)k}\gamma^\mu P_L$, $\frac{ig}{\sqrt{2}}U^{\nu*}_{(a+3)k}\gamma^\mu P_L$\\
		\hline
		$  H_1^+\overline{n_k} l_b$, 	$H_1^-\overline {l_a} n_k $ & $-\frac{ig}{\sqrt{2}m_W}\left(\lambda^{L,1}_{bk}P_L+\lambda^{R,1}_{bk}P_R\right)$,  $-\frac{ig}{\sqrt{2} m_W}\left(\lambda^{L,1*}_{ak}P_R+\lambda^{R,1*}_{ak}P_L\right)$ \\
		\hline
		$  H_2^+\overline{n_k} l_b$, 	$H_2^-\overline {l_a} n_k $ & $-\frac{igc_{23}}{m_W}\left(\lambda^{L,2}_{bk}P_L+\lambda^{R,2}_{bk}P_R\right)$,  $-\frac{igc_{23}}{m_W}\left(\lambda^{L,2*}_{ak}P_R+\lambda^{R,2*}_{ak}P_L\right)$ \\	
		\hline
		$ h^0_1 \overline {l_a}l_a$, $ h^0_2 \overline {l_a}l_a$, $ h^0_3 \overline {l_a}l_a$ &  $-\frac{igm_{l_a}}{2m_W}c_\alpha$, $-\frac{igm_{l_a}}{2m_W}s_\alpha$, $-\frac{igm_{l_a}}{2m_W}$  \\
		\hline
		$ h^0_1 \overline {n_k}n_j$, $ h^0_2 \overline {n_k}n_j$, $ h^0_3 \overline {n_k}n_j$ &   $\frac{igc_{\alpha}}{m_W}\left(\lambda^{h^0_1}_{kj}P_L+\lambda^{h^0_1*}_{kj}P_R\right)$, $-\frac{igs_{\alpha}}{m_W}\left(\lambda^{h^0_2}_{kj}P_L+\lambda^{h^0_2*}_{kj}P_R\right)$, $-\frac{ig}{m_W}\left(\lambda^{h^0_3}_{kj}P_L+\lambda^{h^0_3*}_{kj}P_R\right)$ \\
		\hline
		$h^0_1 H^+_1H^-_1$ & $s_{\alpha } v_3 \left(f+\lambda _{12}\right)-\sqrt{2} v c_{\alpha} \lambda _1 $ \\
		\hline
		$ h^0_1H^+_2H^-_2$ & $\left(s_{\alpha} v_3 \lambda _{12}-\sqrt{2} v c_{\alpha} \lambda _1\right) c_{13}^2-\sqrt{2} f c_{\alpha}
		s_{13} v_3 c_{13}+\frac{1}{2} s_{13}^2 \left(4 s_{\alpha} v_3 \lambda _2-\sqrt{2} v c_{\alpha} \lambda _{12}\right)$ \\
		\hline
		$V_{\mu}^-H_2^+h^0_1$, $V^+_{\mu}H_2^-h^0_1$ & $-\frac{1}{4} ig  \left(\sqrt{2} c_{23} c_{\alpha }+2 s_{23} s_{\alpha }\right)(p^0_1-p_2^+)^\mu, \frac{1}{4} i g
		 \left(\sqrt{2} c_{23} c_{\alpha }+2 s_{23} s_{\alpha }\right)(p^0_1-p_2^-)^\mu $ \\
		\hline
		$h^0_1 W^+_{\mu }W^-_{\nu}$, $h^0_2 W^+_{\mu }W^-_{\nu}$ & $-\frac{gm_Wc_{\alpha }}{\sqrt{2}} g^{\mu \nu}$, $\frac{gm_Ws_{\alpha }}{\sqrt{2}}g^{\mu \nu}$ \\
		\hline
		$h^0_1 V^+_{\mu }V^-_{\nu}$, $h^0_2 V^+_{\mu }V^-_{\nu}$, $h^0_3 V^+_{\mu }V^-_{\nu}$ &  $-\frac{1}{4} gm_W \left(\sqrt{2}  c_{\alpha }-
		\frac{2s_{\alpha }}{t_{23}}\right)g^{\mu \nu}$, $\frac{1}{4} gm_W \left(\sqrt{2}  s_{\alpha }+\frac{2c_{\alpha}}{t_{23}})g^{\mu \nu}\right) $, $-\frac{gm_W}{2 \sqrt{2}}g^{\mu \nu}$ \\
		\hline
		$ h^0_2H^+_1H^-_1$ & $\sqrt{2} v s_{\alpha } \lambda _1+c_{\alpha } v_3 \left(f+\lambda _{12}\right) $\\
		\hline
		$h^0_2 H^+_2H^-_2$ & $\left(\sqrt{2} v s_{\alpha } \lambda _1+c_{\alpha } v_3 \lambda _{12}\right) c_{13}^2+\sqrt{2} f s_{13}
		s_{\alpha} v_3 c_{13}+\frac{1}{2} s_{13}^2 \left(4 c_{\alpha } v_3 \lambda _2+\sqrt{2} v s_{\alpha } \lambda _{12}\right)$ \\
		\hline
		$V_{\mu}^-H_2^+h^0_2$, $V^+_{\mu}H_2^-h^0_2$ & $-\frac{1}{4} i g \left(2 s_{13} c_{\alpha }-\sqrt{2} c_{13} s_{\alpha }\right)(p^0_2-p_2^+)^\mu ,\frac{1}{4} i g
		 \left(2 s_{13} c_{\alpha }-\sqrt{2} c_{13} s_{\alpha }\right)(p^0_2-p_2^-)^\mu $  \\
		\hline
		$h^0_3 H^+_2H^-_2$ & $\sqrt{2}fc_{13} s_{13} v_3$ \\
		\hline
		$V_{\mu}^-H_2^+h^0_3$, $V^+_{\mu}H_2^-h^0_3$ & $-\frac{ig}{2 \sqrt{2}}c_{13}(p^0_3-p_2^+)^\mu,\frac{ig}{2 \sqrt{2}}c_{13}(p^0_3-p_2^-)^\mu $ \\
		\hline
		$W_{\mu}^-H_1^+h^0_3$, $W^+_{\mu}H_1^-h^0_3$ & $ -\frac{1}{2} i g (p^0_3-p_1^+)^\mu , \frac{1}{2} i g (p^0_3-p_1^-)^\mu $ \\
		\hline
		$ h^0_4H^+_1H^-_2$ & $-\sqrt{2} f c_{13} s_{13} v_3$ \\
		\hline
		$V_{\mu}^-H_1^+h^0_4$, $V^+_{\mu}H_1^-h^0_4$ &  $-\frac{i c_{13} g }{2 \sqrt{2}}(p^0_4-p_1^+)^\mu, -\frac{i c_{13} g}{2 \sqrt{2}}(p^0_4-p_1^-)^\mu $\\
		\hline
		$W_{\mu}^-H_2^+h^0_4$, $W^+_{\mu}H_2^-h^0_4$ & $-\frac{1}{2} i g s_{13}^2 (p^0_4-p_2^+)^\mu, \frac{1}{2} i g s_{13}^2 (p^0_4-p_2^-)^\mu$ \\
		\hline
	\end{tabular}
	\caption{Couplings related to the SM-like Higgs decay ($\mathrm{H}\rightarrow l_al_b$) in the 331ISS model. All momenta in the Feynman rules corresponding  to these vertices are incoming.  \label{numbers}}
\end{table}

Based on the couplings in Tab.\ref{numbers}, we obtain the one-loop Feynman diagrams contributing to amplitude of $H\rightarrow l_a^{\pm}l_b^{\mp}$, $H \equiv h_1^0, h^0_2, h^0_3, h^0_4$ in the unitary gauge, which are shown in Fig.\ref{fig_hmt}. All Feynman diagrams are given follow:
\begin{figure}[ht]
	\centering
	\begin{tabular}{cc}
		\includegraphics[width=15.0cm]{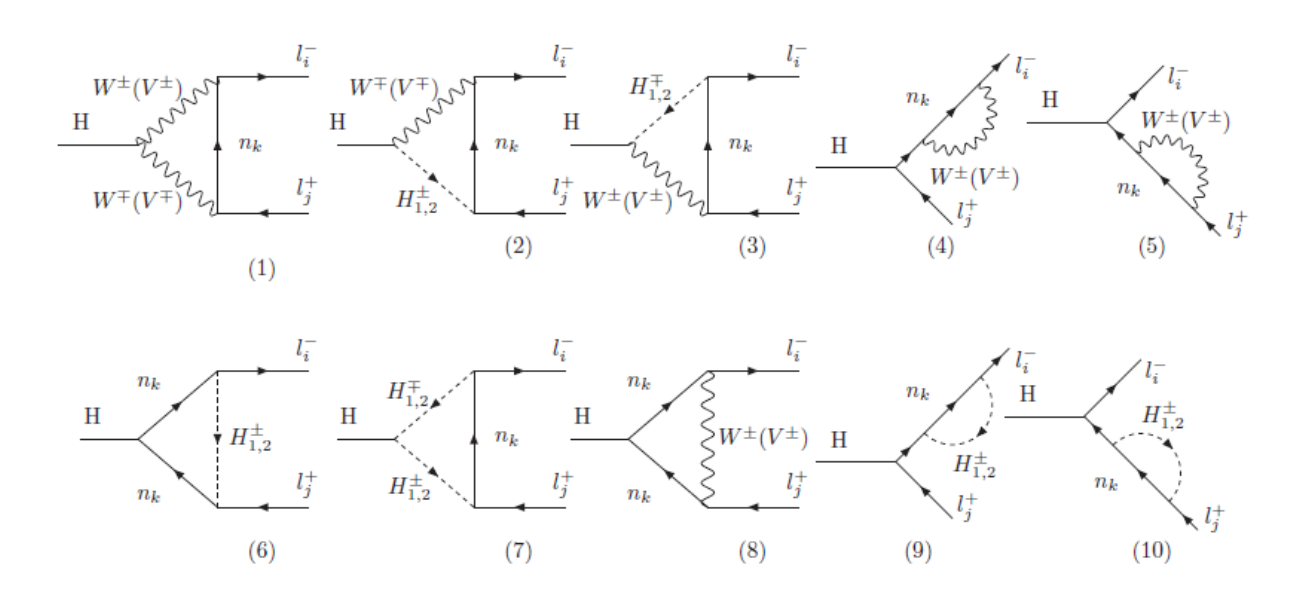} 
	\end{tabular}%
	\caption{ Feynman diagrams at one-loop order of $\mathrm{H} \rightarrow l_il_j$ decays in unitary gauge.}
	\label{fig_hmt}
\end{figure}

We use scalar factors $\Delta_{(ab)L}$ and $\Delta_{(ab)R}$ quoted from 
$ \mathcal{L}_{\mathrm{LFVH}}^\mathrm{eff}= H \left(\Delta_{(ab)L} \overline{l_a}P_L l_b +\Delta_{(ab)R} \overline{l_a}P_R l_b\right) + \mathrm{h.c.},$
as represented as Ref.\cite{Nguyen:2018rlb, Hung:2021fzb}.\\
The partial width of $H\rightarrow l_a^{\pm}l_b^{\mp}$ is
\be
\Gamma (H\rightarrow l_al_b)\equiv\Gamma (H\rightarrow l_a^{+} l_b^{-})+\Gamma (H\rightarrow l_a^{-} l_b^{+})
=  \fr{ m_{H} }{8\pi }\left(\vert \Delta_{(ab)L}\vert^2+\vert \Delta_{(ab)R}\vert^2\right), \label{LFVwidth}
\ee
With conditions $p^2_{1,2}=m^2_{a,b}$,\, $(p_1+p_2)^2=m^2_{H}$ and $m^2_{H}\gg m^2_{a,b}$,\, we obtain branching ratio is $Br(H\rightarrow l_al_b)=\Gamma (H\rightarrow l_al_b)/\Gamma^\mathrm{total}_{H}$, where $\Gamma^\mathrm{total}_{H}\simeq 4.1\times 10^{-3}\mathrm{GeV}$ corresponding to $m_{H}\equiv m_{h^0_1}=125.09\,[\mathrm{GeV}]$ as shown in Refs. \cite{Patrignani:2016xqp,Zyla:2020zbs,CMS:2022ahq,Denner:2011mq}. We also use the current constraints $Br(H\rightarrow \mu\tau) < 10^{-3}$ from Ref.\cite{Aad:2012tfa, Chatrchyan:2012ufa}, so the upper bound for $\Gamma (H \rightarrow \mu\tau)$ is : $\Gamma(H\rightarrow \mu\tau)< 4.1 \times 10^{-6} GeV$.

Similarly in Ref.\cite{Hung:2022vqx}, we can give a coefficients,
\bea \Delta_H^2(H \rightarrow l_al_b)=\Delta_L^2+\Delta_R^2=8\pi \times \frac{\mathrm{\Gamma}(\mathrm{H} \rightarrow l_al_b)}{m_{\mathrm{H}}}\sim \frac{\mathrm{\Gamma}(\mathrm{H} \rightarrow l_al_b)}{m_{\mathrm{H}}}. \label{delta2}
\eea
This is a quantity that is proportional to the partial width ($\mathrm{\Gamma}(\mathrm{H} \rightarrow \mu\tau)$) per unit mass of a particle and completely dependent on the diagrams in Fig.\ref{fig_hmt}.\\
From Eqs.(\ref{LFVwidth},\ref{delta2}), we have:
\bea
m_H=m_{h^0_1} \times \frac{\Gamma(H \rightarrow l_al_b)}{\Gamma(h^0_1 \rightarrow l_al_b)} \times \frac{\Delta^2_{h^0_1}(h^0_1 \rightarrow l_al_b)}{\Delta^2_{H}(H \rightarrow l_al_b)}, \label{mH}
\eea
Based on this relationship, we will predict the masses of the CP-even Higgs bosons in the numerical investigation below.

\section{\label{Analytic} Contribution components to Amplitude of $H \rightarrow l_al_b$} 
In this section, we will give analytic formulas for the components that contribute to the amplitude of $h^0_1 \rightarrow l_al_b$. The divergence parts in specific contributions are indicated. In addition, we also show that the divergences in the total amplitude have zero sums. For other cases of CP-even Higgs bosons are performed similarly and are shown in appendix \ref{appen_loops2}.\\
From Tab.(\ref{numbers}), we can see that all diagrams in Fig.(\ref{fig_hmt}) participate in decay of $h^0_1 \rightarrow l_al_b$. However, the results show that $h^0_1W^\pm H_s^\mp =0$ so the 2nd and 3rd diagrams in Fig.(\ref{fig_hmt}) only give the contribution of the $V$-boson. 
\bea \Delta_{(ab)L,R}^{h^0_1} &&= c_\alpha \times \mathcal{A}^{(1)W}_{L,R}\left( m_W\right)  +\frac{\left(\sqrt{2}s_\alpha c_{23} -c_\alpha s_{23} \right)}{\sqrt{2}} \times \mathcal{A}^{(1)V}_{L,R}\left( m_V\right) \crn
&& +\frac{\left(\sqrt{2}s_\alpha s_{23} +c_\alpha c_{23} \right)}{\sqrt{2}t_{23}} \times \mathcal{A}^{(2+3)V}_{L,R}\left( m_V,m_{H^\pm_2}\right)  +c_{\alpha}\times \mathcal{A}^{(4+5)W}_{L,R}\left( m_{W}\right)  \crn
&&+\frac{c_{\alpha}}{\sqrt{2}s_{23}}\times \mathcal{A}^{(4+5)V}_{L,R}\left( m_{V}\right)-c_{\alpha}\times\mathcal{A}^{(6)H^\pm_1}_{L,R}\left( m_{H^\pm_1}\right)-2c_{\alpha}c_{23}^2\times \mathcal{A}^{(6)H^\pm_2}_{L,R}\left( m_{H^\pm_2}\right)  \crn
&&+ 2\times \mathcal{A}^{(7)H^\pm_1}_{L,R}\left( m_{H^\pm_1}\right) + 4c^2_{23}\times \mathcal{A}^{(7)H^\pm_2}_{L,R}\left( m_{H^\pm_2}\right) + c_{\alpha}\times\mathcal{A}^{(8)W}_{L,R}\left( m_{W}\right) \crn
&& +\frac{c_{\alpha}}{\sqrt{2}s_{23}}\times \mathcal{A}^{(8)V}_{L,R}\left( m_{V}\right)+ c_{\alpha}\times \mathcal{A}^{(9+10)H^\pm_1}_{L,R}\left( m_{H^\pm_1}\right) + 2c_{\alpha}c_{23}^2 \times \mathcal{A}^{(9+10)H^\pm_2}_{L,R}\left( m_{H^\pm_2}\right),  \label{deLR1}
\eea
Next, we pay attention to the divergences contained in the amplitudes as given in App.\ref{appen_loops1}. Using the divergence separation of PV-functions such as ref.\cite{Nguyen:2018rlb, Hung:2021fzb}: div$B^{(1)}_0=$div$B^{(2)}_0=$div$B^{(12)}_0=2$div$B^{(1)}_1=-2$ div$B^{(2)}_1=\Delta_{\epsilon}$, replacing $1/m_V=\sqrt{2}s_{23}/m_W$ and omitting the common term $g^3/(64\pi^2m_G^3)$ in the expression, we represent the divergence as follows :  
\bea 
&&\mathrm{div}\left[\mathcal{A}^{(1)G}_R\right]=m_b\Delta_{\epsilon}\times \left( -\frac{3}{2}\right) \sum_{i=1}^9 U^{\nu*}_{G}U^{\nu}_{G} m^2_{n_i},\,
\mathrm{div}\left[\mathcal{A}^{(2)V}_R\right]=-\Delta_{\epsilon}\times \sum_{i=1}^9 U^{\nu*}_{(a+3)i}\lambda^{R,s}_{bi} m_{n_i}^2,\crn
&&\mathrm{div}\left[\mathcal{A}^{(3)V}_R\right]=m_b\Delta_{\epsilon}\times \frac{1}{2}  \sum_{i=1}^9 U^{\nu}_{(b+3)i}\lambda^{L,s*}_{ai} m_{n_i},\,\mathrm{div}\left[\mathcal{A}^{(6)H_s^\pm}_R\right]=\Delta_{\epsilon}\times  \sum_{i,j=1}^9 \lambda^{h^0_1}_{ij}\lambda^{L,s*}_{ai}\lambda^{R,s}_{bj} ,\crn
&&\mathrm{div}\left[\mathcal{A}^{(8)G}_R\right]=m_b \Delta_{\epsilon}\times \sum_{i,j=1}^9 U^{\nu*}_{G}U^{\nu}_{G}\left(\lambda^{h^0_1*}_{ij} m_{n_j}+\frac{1}{2}\lambda^{h^0_1}_{ij} m_{n_i}\right), \crn
&&\mathrm{div}\left[\mathcal{A}^{(4+5)W}_R\right]=\mathrm{div}\left[\mathcal{A}^{(7)H^\pm_s}_R\right]= \mathrm{div}\left[\mathcal{A}^{(4+5)V}_R\right]=0,
\label{divdeli}\eea
We use notation $\Delta^{(k)}_{L,R}$ for the left-right components of the amplitude of the $k$th diagram in Figure \ref{fig_hmt} and note the formulas in App.\ref{CaDV}. The divergences are arranged to cancel each other out as follows:
\begin{align}
	& \mathrm{div}\left[\Delta^{(1)W}_R\right]+\mathrm{div}\left[\Delta^{(8)W}_R\right]=(m_D^\dagger m_D)_{ba} \left( -\frac{3}{2}c_\alpha +\frac{3}{2}c_\alpha  \right) = 0\crn
	&\mathrm{div}\left[\Delta^{(6)H^\pm_1}_R\right]+\mathrm{div}\left[\Delta^{(9+10)H^\pm_1}_R\right]=(m_D^\dagger m_D)_{ba} \left( c_\alpha -c_\alpha  \right)= 0 \crn
	& \mathrm{div}\left[\Delta^{(1+2+3+8)V}_R+\Delta^{(6+9+10)H^\pm_2}_R \right]\crn
	&\sim (m_D^\dagger m_D)_{ba} \left\{\sqrt{2}s_{\alpha}s^2_{23}c_{23}(3-1-2) + c_{\alpha}\left[s^2_{23}(-3s^2_{23}-c^2_{23}-2c^2_{23}+3) +2 s^2_{23}-2s^2_{23} \right]\right\}\crn
	&+ (M^*_RM^T_R)_{ba}\left[ \sqrt{2}s_{\alpha}\frac{s^2_{23}}{c_{23}} \left(3c^2_{23}+s^2_{23}-2c^2_{23}-3+2\right)+ c_{\alpha}s^2_{23}\left(-3s^2_{23}+s^2_{23}-2c^2_{23}+2\right)\right]\crn
	&=0. \label{sdiv}
\end{align}
By the same way, we can show that the divergence is also eliminated in the left component ($\Delta_L$) of the total amplitude.

\section{\label{Numerical} Numerical results} 
\subsection{Setup parameters}  
We use the well-known experimental parameters \cite{Patrignani:2016xqp}: 
the charged lepton masses $m_e=5\times 10^{-4}\,\mathrm{GeV}$,\,  $m_\mu=0.105\,\mathrm{GeV}$,\, $m_\tau=1.776\,\mathrm{GeV}$,\, the SM-like Higgs mass $m_{h^0_1}=125.1\,\mathrm{GeV}$,\,  the mass of the W boson $m_W=80.385\,\mathrm{GeV}$  and the gauge coupling of the $SU(2)_L$ symmetry $g \simeq 0.651$.\\
For convenience, we choose a set of free
parameters  are: mass of charged gauge boson $m_V$, Higgs self-coupling constants $\lambda_1,\, \lambda_{12}$, mass of charged Higgs $m_{H^\pm_1}$, $\varrho$ to numerically investigate of both decays $l_a \rightarrow l_b\gamma$ and $H \rightarrow l_al_b$. Therefore, the dependent parameters are given follows.
\bea 
v&=\frac{\sqrt{2}m_W}{g}, \, s_{23}=\frac{m_W}{\sqrt{2}m_V},\,v_3=\frac{2m_V}{gc_{23}},\crn
f&=\frac{gc_{23} m^2_{H^\pm_1}}{4m_V},\, m^2_{H^\pm_2}=\frac{m^2_{H^\pm_1}}{2}\left(t^2_{23} +1 \right) ,\crn
\lambda_2&=\frac{t^2_{23} }{2}\left(\frac{2m^2_{h^0_1}}{v^2}-\frac{m^2_{H^\pm_1}}{2v_3^2} \right) +\frac{\left( \lambda_{12}-\frac{m^2_{H^\pm_1}}{2v_3^2}\right) ^2}{4\lambda_1-\frac{2m^2_{h^0_1}}{v^2}} .\label{deppara}
\eea 
As results in Refs.\cite{CMS:2018ipm,ATLAS:2019erb}, we can give a limit to the mass of the new charged gauge boson ($m_V$) based on condition $m_Z'\geq 4.0\,\mathrm{TeV}$. In this model, we have $m_Z'^2=\frac{g^2v_3^2c_W^2}{3-4s_W^2}$ resulting in $v_3\geq 10.1\, \mathrm{TeV}$. So, we can keep the value fixed $m_V=4.5~\mathrm{TeV}$ at LHC$@13\mathrm{TeV}$. This value of $m_V$ is very suitable and will be shown in the numerical investigation below.

 The values of  the Higgs self-couplings are chosen to satisfy the currently known constraints $\lambda_1=1,\, \lambda_{12}=-1$. Based on recent data of neutral meson
mixing $B_0 - \overline{B}_ 0$  \cite{Okada:2016whh}, we can choose the lower bound of $m_{H^\pm_1}\geq 500\mathrm{GeV}$. This is also consistent with Refs.\cite{Nguyen:2018rlb, Hung:2021fzb}. The parameter $\varrho$ of the matrix $m_D$ in Eq.(\ref{nmD}) is considered in the range of the perturbative limit, $\varrho =\sqrt{2}v_2h^\nu_{23}\leq 617 \,\mathrm{GeV}$. Therefore, to avoid boundary effects, we usual fix the value for $\varrho$ about $500\, \mathrm{GeV}$ in the calculations below. To represent masses of heavy neutrinos ($F_a$), we parameterize the matrix $M_R$ in the form of a diagonal. More interestingly, we can also exploit some consequences from the hierarchy of a diagonal matrix $M_R$. 

\subsection{ Numerical results of $H \rightarrow \mu \tau$ decays}   
\begin{figure}[ht]
	\centering
	\begin{tabular}{c}
		\includegraphics[width=10.0cm]{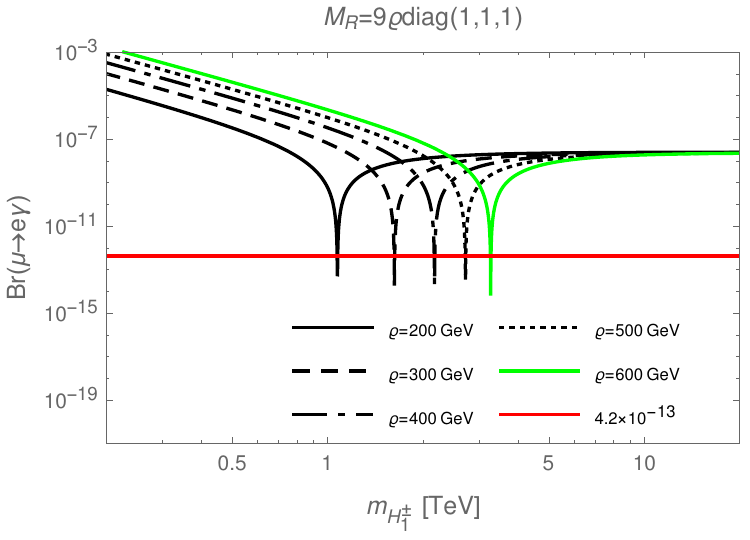} 
	\end{tabular}%
	\caption{ Plots of $Br(\mu \rightarrow e \gamma)$ depend on $m_{H^\pm_1}$ in case $M_R=9\varrho diag(1,1,1)$.}
	\label{fig_lepton1}
\end{figure}

In this section, we will investigate $\Gamma(H \rightarrow \mu \tau)$ in regions of parameter space that satisfy the experimental limit of charged leptons decays $l_a \rightarrow l_b \gamma $. Depending on the form of the mixing matrix of the heavy neutrinos ($M_R$), we have different regions of the allowed parameter space. Using the results as shown in Refs.\cite{Nguyen:2018rlb, Hung:2021fzb}, we will choose $M_R$ with diagonal form and parameterized according to the cases: $M_R=9\varrho diag(1,1,1)$, $M_R=9\varrho diag(1,2,3)$, $M_R=9\varrho diag(3,2,1)$. The hierarchy of $M_R$ in the experimental constraints of $\Gamma(H \rightarrow \mu \tau)$  and $l_a \rightarrow l_b \gamma $ entails a change in the ratio of the mass between the SM-like Higgs boson and the other CP-even Higgs bosons. We are concerned with the case of $M_R=9\varrho diag(1,1,1)$ first, the remaining cases will be considered similarly. As results in Ref.\cite{Hung:2021fzb}, the interference between the contributing components of the charged Higgs bosons and the charged gauge bosons produces narrow parameter space regions satisfying $Br(\mu \rightarrow e \gamma)< 4.2 \times 10^{-13}$  as shown in Fig.\ref{fig_lepton1}. 

\begin{figure}[ht]
	\centering
	\begin{tabular}{cc}
		\includegraphics[width=8.0cm]{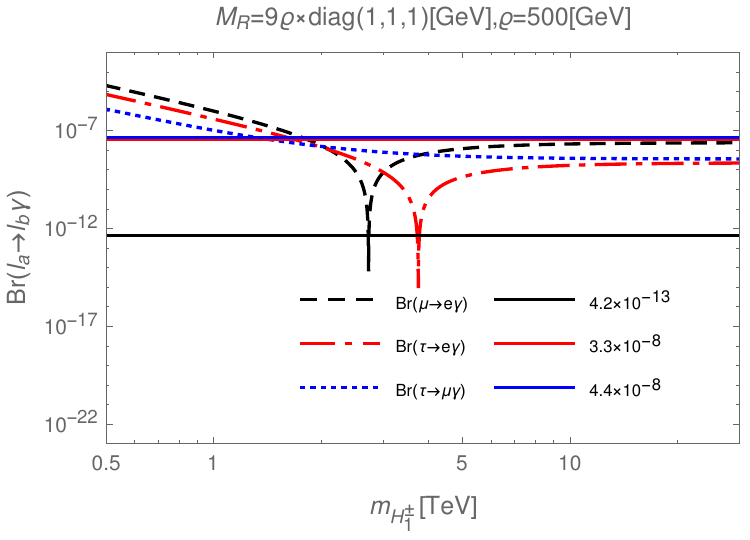} & \includegraphics[width=5.5cm]{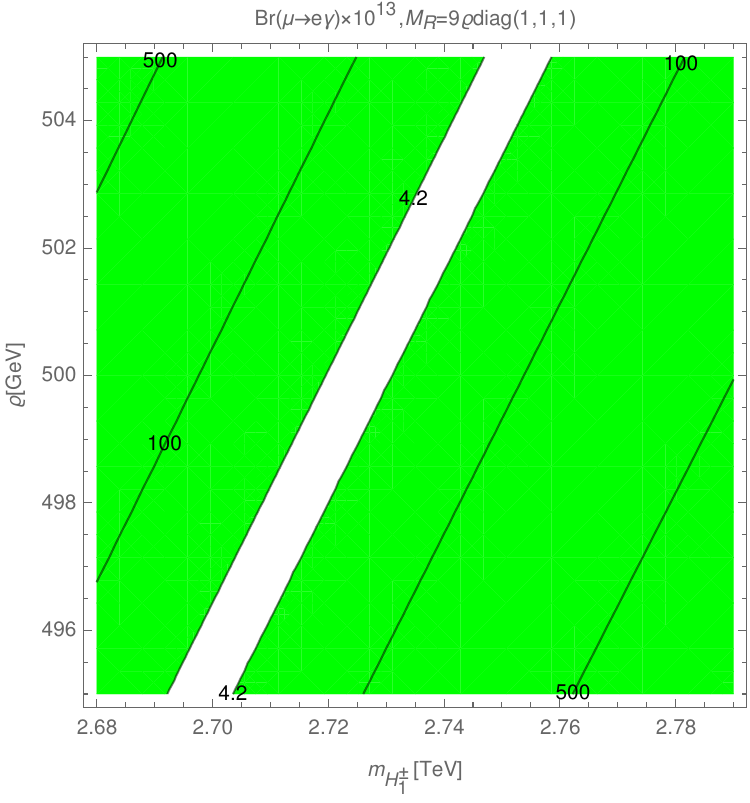} 
	\end{tabular}%
	\caption{ Plots of $Br(l_a \rightarrow l_b \gamma)$ and contour plots of $Br(\mu \rightarrow e\gamma)$ depend on $m_{H^\pm_1}$ in case $M_R=9\varrho diag(1,1,1)$.}
	\label{fig_lepton2}
\end{figure}

The $\varrho$ parameter derived from Eq.\ref{nmD} must satisfy the limit of perturbation theory $\varrho =\sqrt{2}v_2h^\nu_{23}\leq 617 \,\mathrm{GeV}$ and as above comment, we will fix $\varrho = 500\, GeV$. The spatial regions satisfy the constraint of $l_a \rightarrow l_b \gamma $ as shown in Fig.\ref{fig_lepton2}. On the left panel of Fig.\ref{fig_lepton2}, it is easy to see that in the region of parameter space where $\mu \rightarrow e \gamma $ satisfies the experimental limits, $\tau \rightarrow e \gamma $ and $\tau \rightarrow \mu \gamma $ also satisfy. Therefore, the colorless in the right panel represents the allowed space, satisfying the experimental constraints of $l_a \rightarrow l_b \gamma $. 

\begin{figure}[ht]
	\centering
	\begin{tabular}{cc}
		\includegraphics[width=7.0cm]{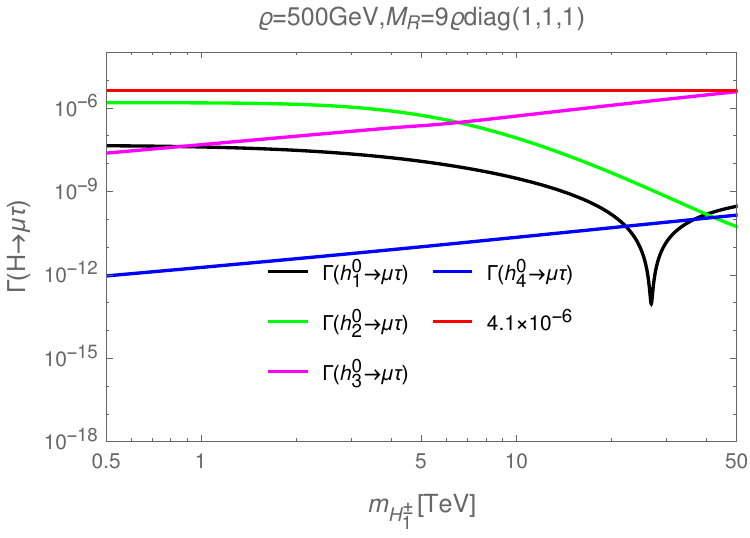} & \includegraphics[width=7.0cm]{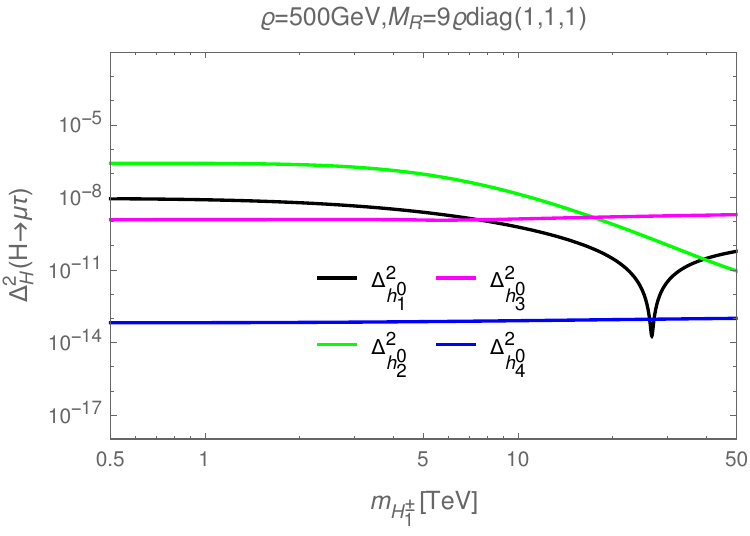} 
	\end{tabular}%
	\caption{ Plots of $\Gamma(H \rightarrow \mu \tau)$ and $\Delta^2_H(H \rightarrow \mu \tau)$ depend on $m_{H^\pm_1}$ in case of $M_R=9\varrho diag(1,1,1)$.}
	\label{fig_Hmt1}
\end{figure}

The dependence of $\Gamma(H \rightarrow \mu \tau)$ and $\Delta^2_H(H \rightarrow \mu \tau)$ on $m_{H^\pm_1}$ are shown in Fig. \ref{fig_Hmt1}. Investigation results show that in the selected parameter space, $\Gamma(H \rightarrow \mu \tau)< 4.1 \times 10^{-6}\, GeV$ for all CP-even Higgs bosons.

\begin{figure}[ht]
	\centering
	\begin{tabular}{cc}
		\includegraphics[width=6.5cm]{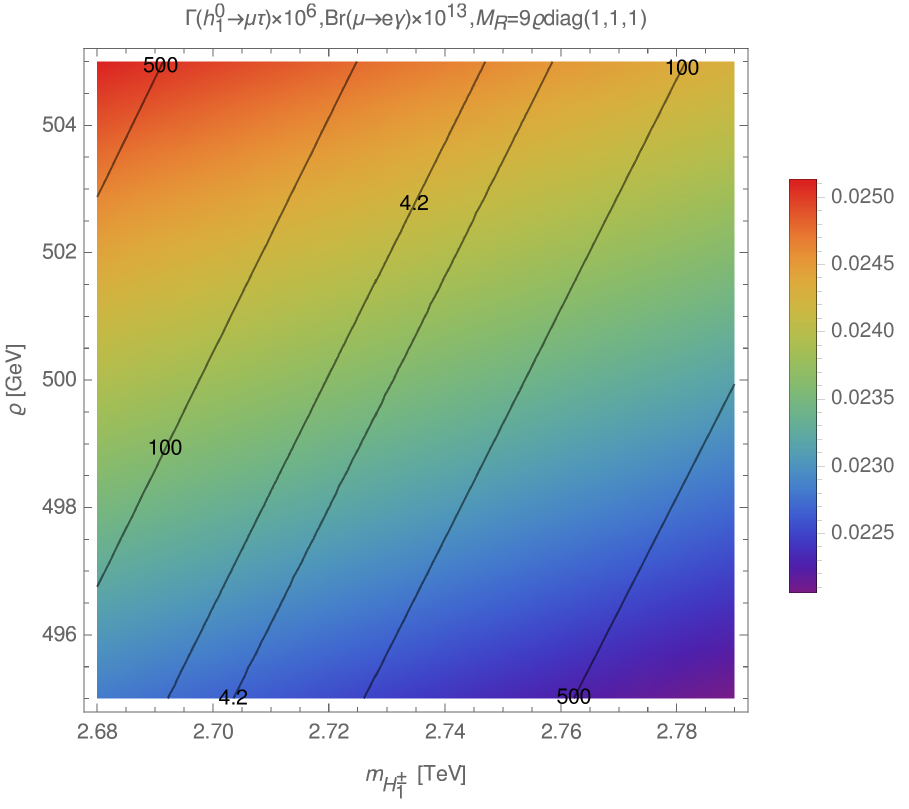} & \includegraphics[width=6.5cm]{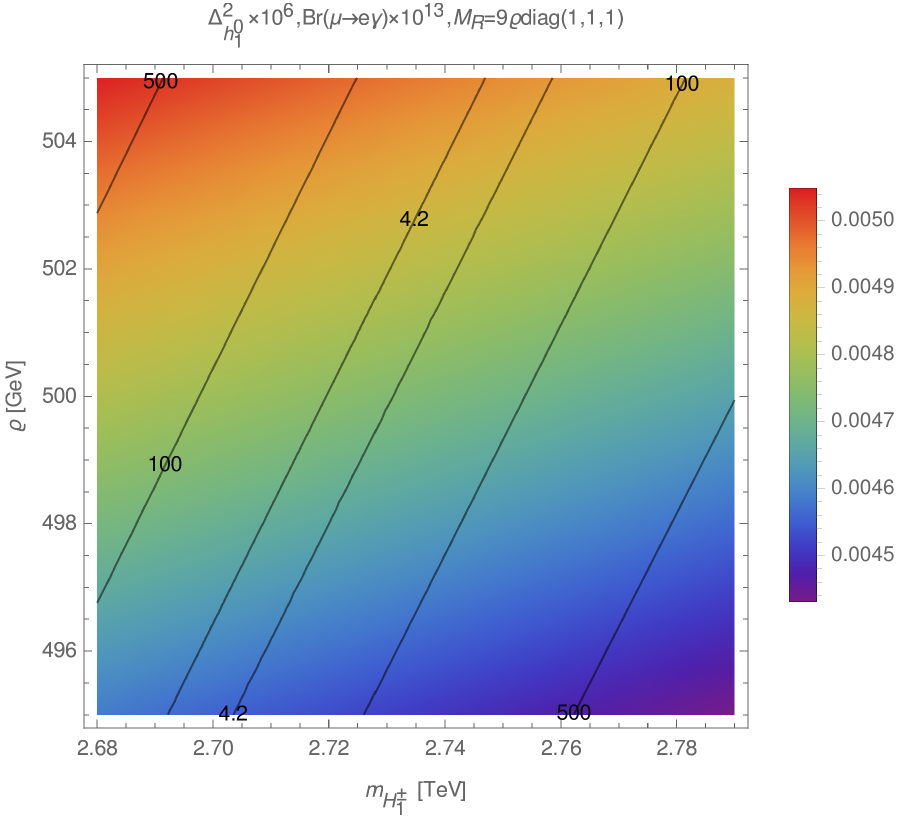}\\ 
		\includegraphics[width=6.5cm]{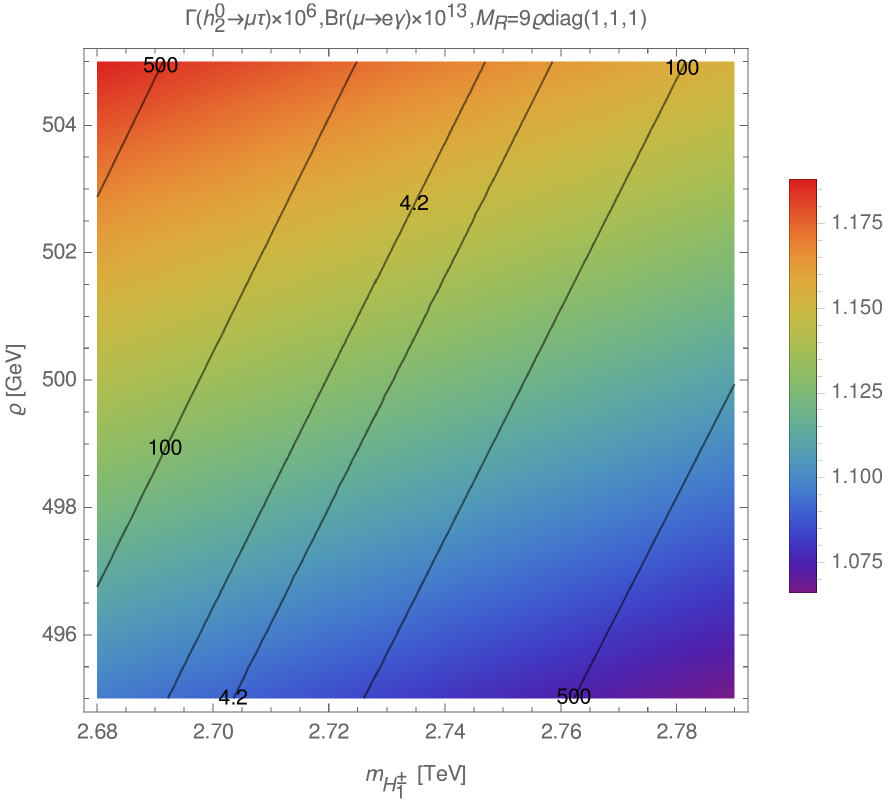} & \includegraphics[width=6.5cm]{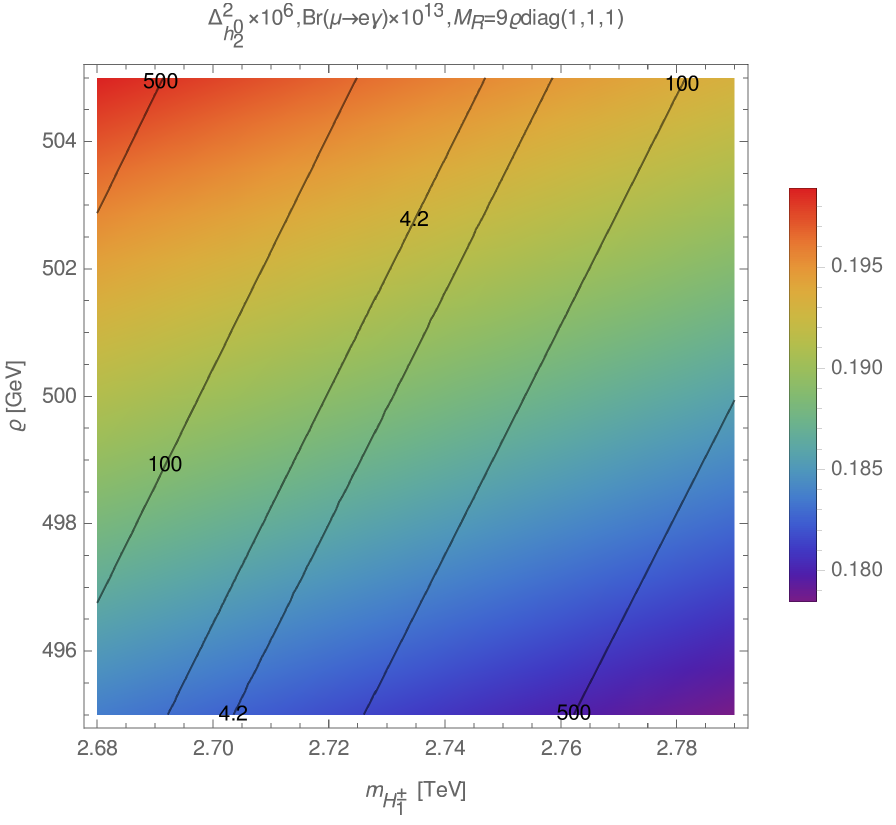}\\
	\end{tabular}%
	\caption{ Density plots of $\Gamma (h^0_1 \rightarrow \mu \tau)$, $\Delta^2_{h^0_1} (h^0_1 \rightarrow \mu \tau)$ (first line) and $\Gamma (h^0_2 \rightarrow \mu \tau)$, $\Delta^2_{h^0_2} (h^0_2 \rightarrow \mu \tau)$ (second line) in the case of $M_R=9\varrho diag(1,1,1)$.}
	\label{fig_Hmt2}
\end{figure}

\begin{figure}[ht]
	\centering
	\begin{tabular}{cc}
		\includegraphics[width=6.5cm]{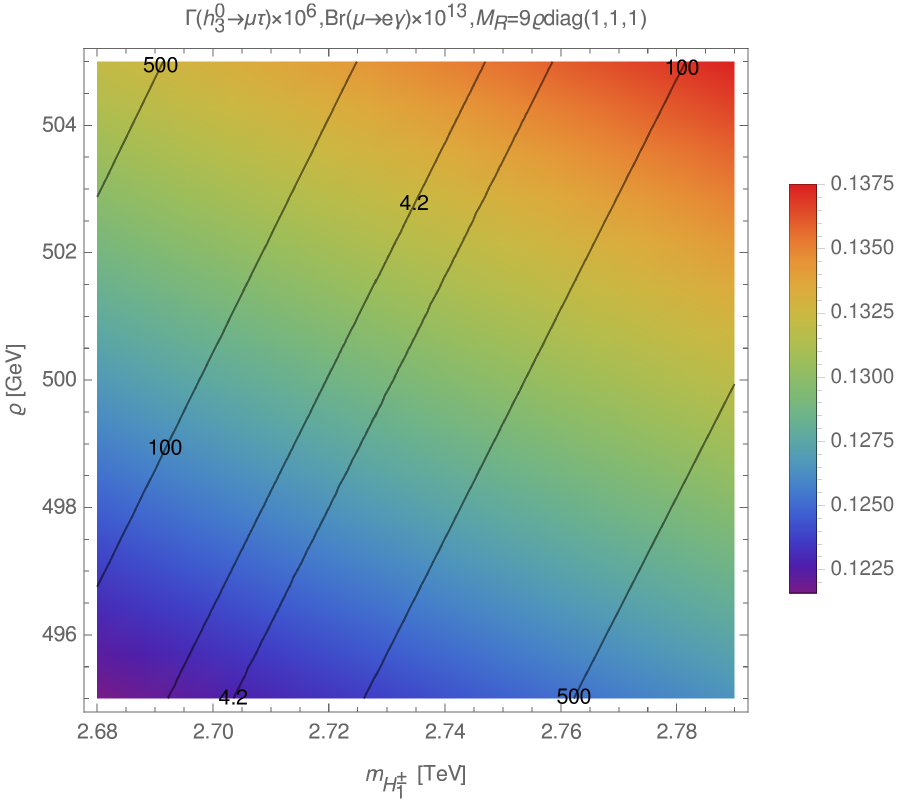} & \includegraphics[width=6.5cm]{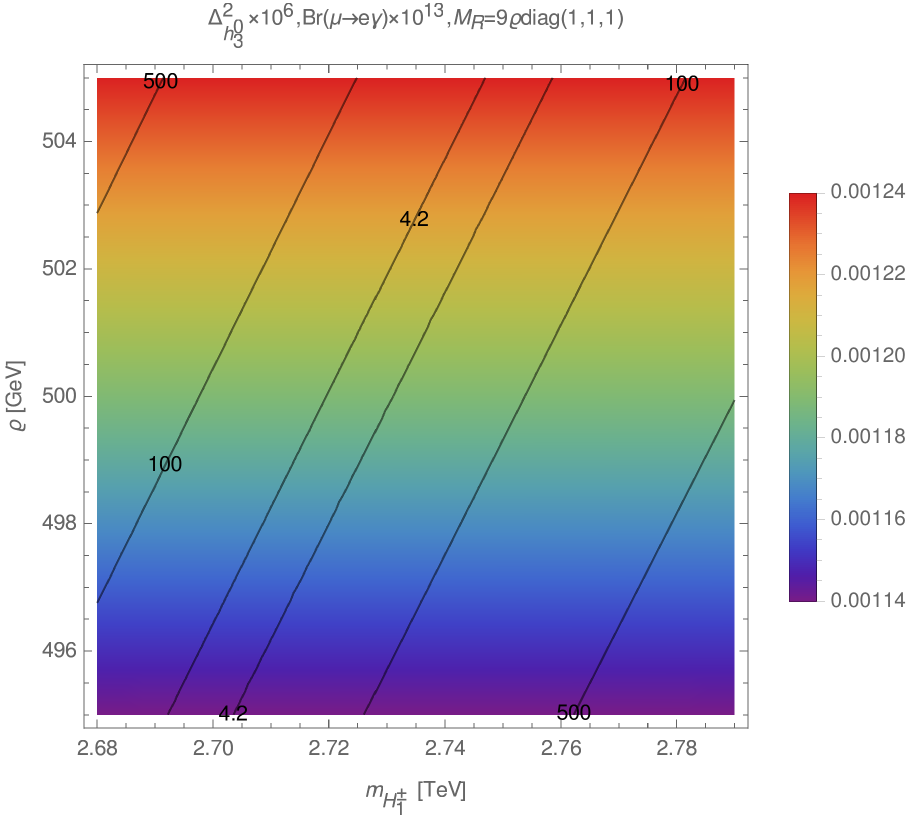}\\ 
		\includegraphics[width=6.5cm]{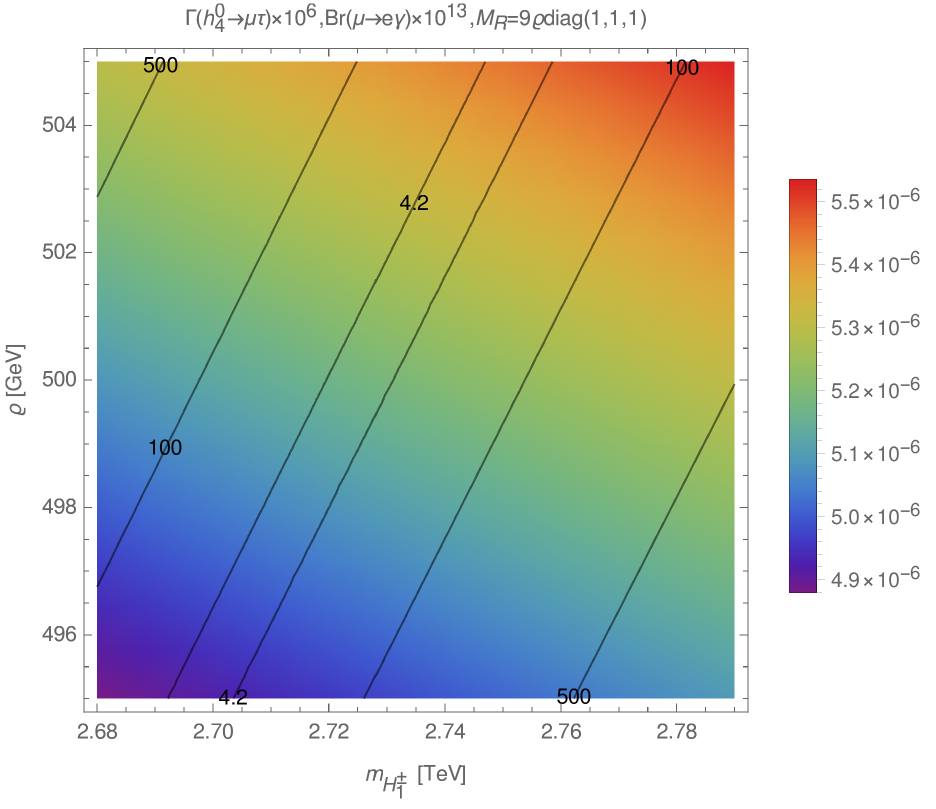} & \includegraphics[width=6.5cm]{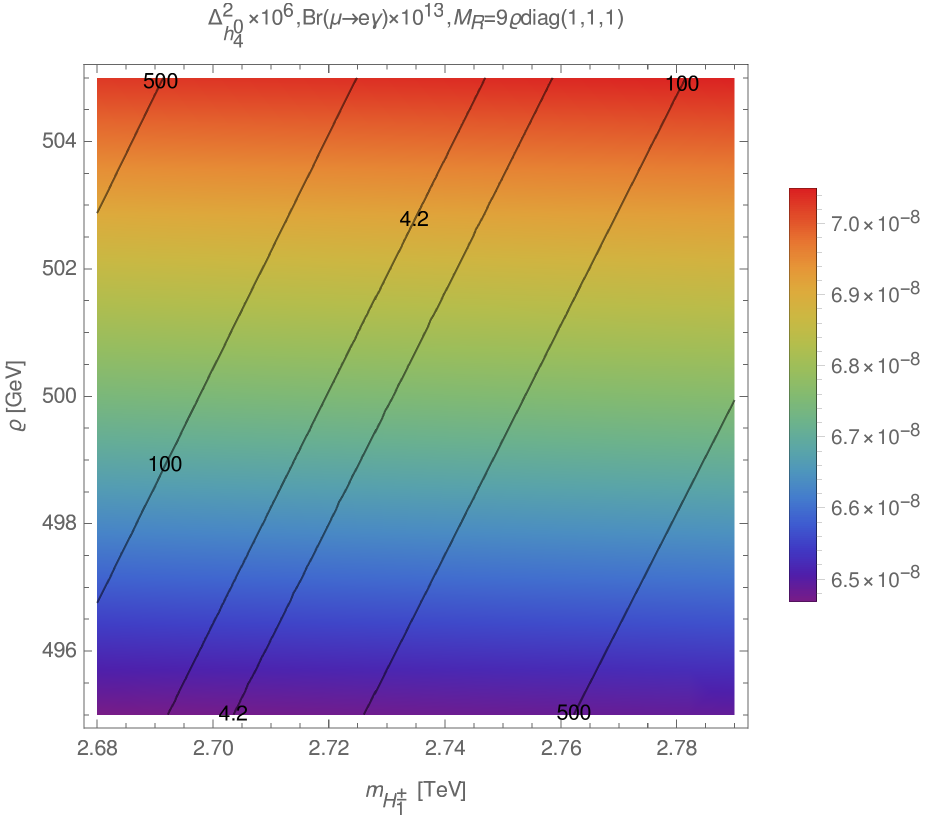}\\ 
	\end{tabular}%
	\caption{ Density plots of $\Gamma (h^0_3 \rightarrow \mu \tau)$, $\Delta^2_{h^0_3} (h^0_3 \rightarrow \mu \tau)$ (first line) and $\Gamma (h^0_4 \rightarrow \mu \tau)$, $\Delta^2_{h^0_4} (h^0_4 \rightarrow \mu \tau)$ (second line) in the case of $M_R=9\varrho diag(1,1,1)$.}
	\label{fig_Hmt3}
\end{figure}

Fig.\ref{fig_Hmt2} and Fig.\ref{fig_Hmt3} give us the results of investigating the values of $\Gamma(H \rightarrow \mu \tau)$ and $\Delta^2_H(H \rightarrow \mu \tau)$ in the region of the parameter space satisfying $l_a \rightarrow l_b \gamma$ of $h^0_1$, $h^0_2$, $h^0_3$, $h^0_4$, respectively. Their allowed values are between the two curves $4.2 \times 10^{-13}$, which is the upper bound of $\mu \rightarrow e \gamma$ decay.

Using Eq.(\ref{mH}), we can compare the mass of the SM-like Higgs boson with that of other CP-even Higgs bosons. For example, we estimate $m_{h^0_2}$ in the case of $M_R=9\varrho diag(1,1,1)$ to be $m_{h^0_4}=125.1 \times \frac{0.0048 \times 10^{-6}}{0.024 \times 10^{-6}}\times \frac{5.3 10^{-6}}{6.8 10^{-8}}=1948 \,GeV$.

\begin{figure}[ht]
	\centering
	\begin{tabular}{cc}
	\includegraphics[width=6.0cm]{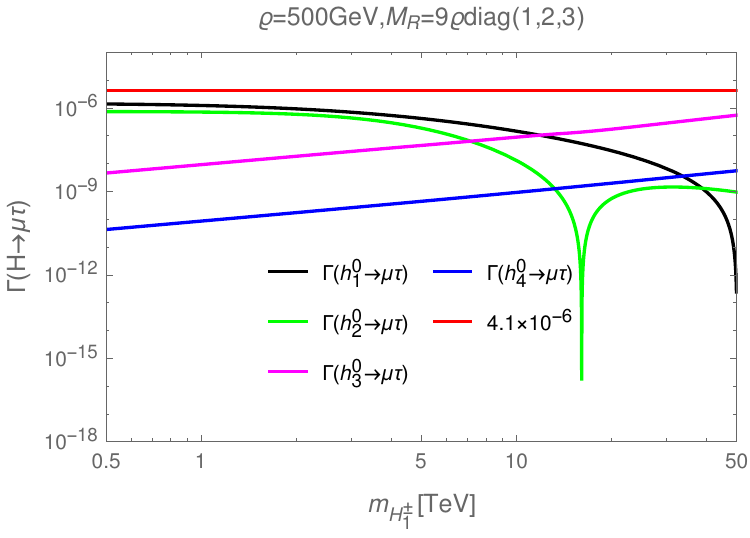} & \includegraphics[width=6.0cm]{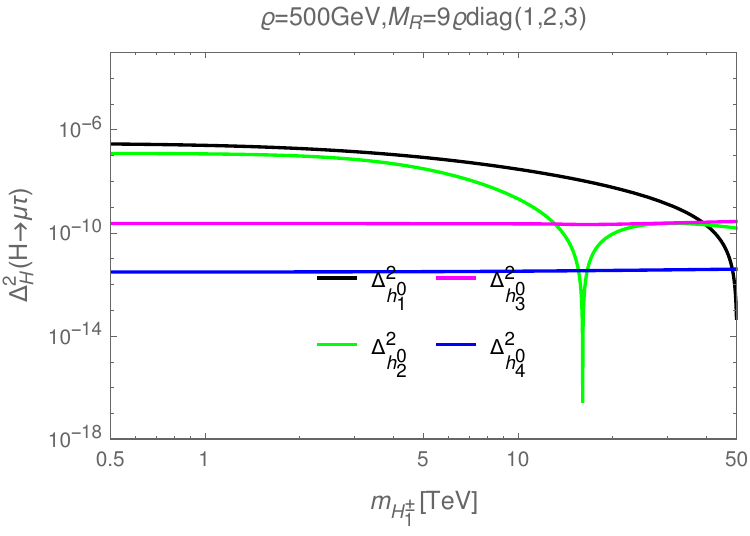}\\ 
	\includegraphics[width=6.0cm]{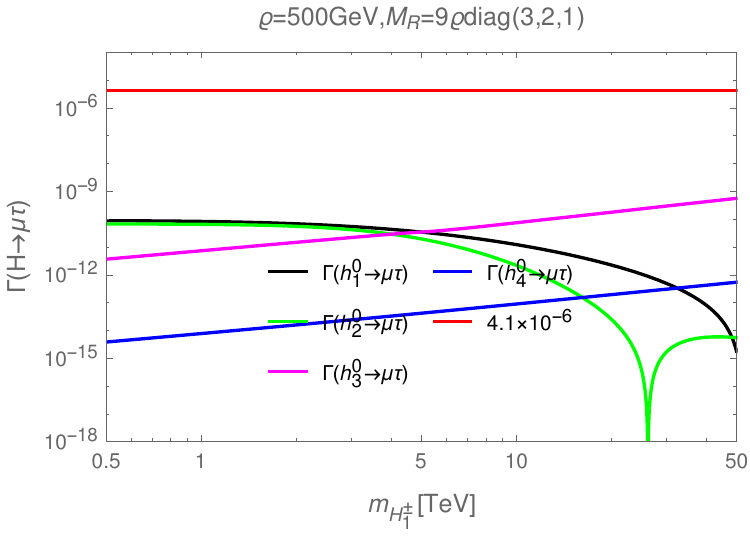} & \includegraphics[width=6.0cm]{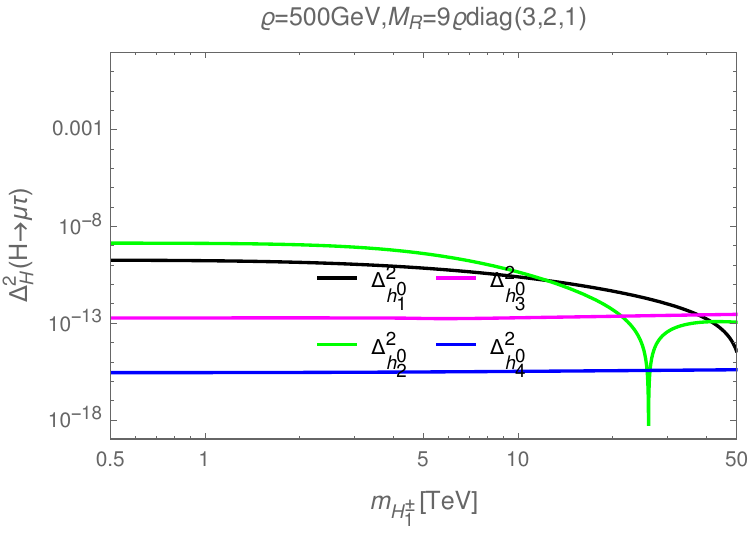}\\ 
	\end{tabular}%
	\caption{ Plots of $\Gamma(H \rightarrow \mu \tau)$ and $\Delta^2_H(H \rightarrow \mu \tau)$ depend on $m_{H^\pm_1}$ in case of $M_R=9\varrho diag(1,2,3)$ (first line) and in case of $M_R=9\varrho diag(3,2,1)$ (second line).}
	\label{fig_Hmt4}
\end{figure}

In a similar way, we also represent $\Gamma (h^0_3 \rightarrow \mu \tau)$ and $\Delta^2_{h^0_3} (h^0_3 \rightarrow \mu \tau)$ to depend on $m_{H^\pm_1}$ for the cases $M_R=9\varrho diag(1,2,3)$ and $M_R=9\varrho diag(3,2,1)$. The results are shown in Fig.\ref{fig_Hmt4}.
Density plots of $\Gamma(H \rightarrow \mu \tau)$ and $\Delta^2_H(H \rightarrow \mu \tau)$  in cases of $M_R=9\varrho diag(3,2,1)$ and $M_R=9\varrho diag(1,2,3)$ are given in App. \ref{appen_loops3}. Based on these results, we also estimate the masses of the other CP-even Higgs bosons as Tab.\ref{values}.
\small{\small{\begin{table}[ht]
	\begin{tabular}{|c|c|c|c|}
		\hline
		 & $M_R=9\varrho diag(1,1,1)$ & $M_R=9\varrho diag(1,2,3)$ & $M_R=9\varrho diag(3,2,1)$ \\
		\hline
	 $\Gamma (h^0_1 \rightarrow \mu \tau)$ & $0.024 \times 10^{-6}$ & $0.29 \times 10^{-6}$ & $0.51 \times 10^{-6}$  \\
		\hline
	$\Delta^2 (h^0_1 \rightarrow \mu \tau)$ & $0.0048 \times 10^{-6}$ & $0.058 \times 10^{-6}$ & $0.102 \times 10^{-6}$  \\
	\hline
	$\Gamma (h^0_2 \rightarrow \mu \tau)$ & $1.125 \times 10^{-6}$ & $0.085 \times 10^{-6}$ & $0.0039 \times 10^{-6}$ \\
	\hline
	$\Delta^2 (h^0_2 \rightarrow \mu \tau)$ & $0.19 \times 10^{-6}$ & $0.14 \times 10^{-6}$ & $0.00074 \times 10^{-6}$\\
	\hline
	$\Gamma (h^0_3 \rightarrow \mu \tau)$ & $0.13 \times 10^{-6}$ & $0.058 \times 10^{-6}$ & $0.0000235 \times 10^{-6}$ \\
	\hline
	$\Delta^2 (h^0_3 \rightarrow \mu \tau)$ & $0.0012 \times 10^{-6}$ & $0.00022 \times 10^{-6}$ & $0.0000001825 \times 10^{-6}$\\
	\hline
	$\Gamma (h^0_4 \rightarrow \mu \tau)$ & $0.0000525 \times 10^{-6}$ & $0.00058 \times 10^{-6}$ & $0.000000026 \times 10^{-6}$ \\
	\hline
	$\Delta^2 (h^0_4 \rightarrow \mu \tau)$ & $0.000000068 \times 10^{-6}$ & $0.00000315 \times 10^{-6}$ & $0.00000000285 \times 10^{-6}$\\
	\hline
	$ m_{h^0_2} \, [GeV]$ & $1420 $ & $1320$ & $1520$\\
	\hline
	$ m_{h^0_3} \, [GeV]$ & $2708 $ & $3219$ & $6590$\\
	\hline
	$ m_{h^0_4} \, [GeV]$ & $1948$ & $2280$ & $4603$\\
	\hline
	\end{tabular}
	\caption{The possible values of $\Gamma(H \rightarrow \mu \tau)$, $\Delta^2_H(H \rightarrow \mu \tau)$ and $m_H$ in the parameter space regions satisfy the experimental limits of $ l_a \rightarrow l_b \gamma$.  \label{values}}
\end{table}}}

In the parameter space region satisfying the experimental limits of $ l_a \rightarrow l_b \gamma$, the numerical investigation results show that the lightest Higgs boson is identified with the SM-like Higgs boson, three other CP-even Higgs bosons with large masses are at the electroweak symmetry breaking scale. This numerical investigation is also consistent with Eqs.(\ref{mixh04},\, \ref{massesnHiggs}). 
 \section{\label{conclusion} Conclusion}
In the 3-3-1 model with inverse seesaw neutrinos, the lepton-flavor-violation couplings depend strongly on the mixing matrix of heavy neutrinos ($M_R$), which creates a great influence on the parameter space regions satisfying the experimental limits of $ l_a \rightarrow l_b \gamma$. We use these regions of parameter space to investigate $\Gamma(H \rightarrow \mu \tau)$ and $\Delta^2_H(H \rightarrow \mu \tau)$.\\
We investigate $\Gamma(H \rightarrow \mu \tau)$ in three typical cases where $M_R$ has a diagonal form: non-hierarchy, normal-hierarchy and inverse-hierarchy. The result obtained $\Gamma(H \rightarrow \mu \tau)$ is always less than the upper limit of the experiment ($4.1 \times 10^{-6}\,GeV)$.\\
We also investigate the $\Delta^2_H(H \rightarrow \mu \tau)$ factors, an important contributor to $\Gamma(H \rightarrow \mu \tau)$, the results obtained helping to predict the masses of other CP-even Higgs bosons: $m_{h^0_2} \in (1320 \div 1520) \, GeV$, $m_{h^0_3} \in (2708 \div 6590) \, GeV$, $m_{h^0_4} \in (1930 \div 4603) \, GeV$. These numerical investigation are consistent with anlytic formulas in Eqs.(\ref{mixh04},\, \ref{massesnHiggs}).

 \section*{Acknowledgments}
 This research is funded by Hanoi Pedagogical University 2 under grant number HPU2.UT-2021.02.
 
 \appendix
 \section{Master integrals.}
 \label{appen_loops}
 To calculate the contributions at one-loop order of the Feynman diagrams in Fig.\ref{fig_hmt}, we use the Passarino-Veltman (PV) functions as mentioned in Ref.\cite{Passarino:1978jh}. By introducing the notations $D_0=k^2-M_0^2+i\delta$, $D_1=(k-p_1)^2-M_{1}^2+i\delta$ and $D_2=(k+p_2)^2-M_2^2+i\delta$, where $\delta$ is  infinitesimally a  positive real quantity, we have:
 \bea
 A_{0}(M_n)
 &\equiv &\frac{\left(2\pi\mu\right)^{4-D}}{i\pi^2}\int \frac{d^D k}{D_n}, \hs
 B^{(1)}_0 \equiv\frac{\left(2\pi\mu\right)^{4-D}}{i\pi^2}\int \frac{d^D k}{D_0D_1},\crn
 B^{(2)}_0 &\equiv &\frac{\left(2\pi\mu\right)^{4-D}}{i\pi^2}\int \frac{d^D k}{D_0D_2}, \hs
 B^{(12)}_0 \equiv \frac{\left(2\pi\mu\right)^{4-D}}{i\pi^2}\int \frac{d^D k}{D_1D_2},\crn
 C_0&\equiv&  C_{0}(M_0,M_1,M_2) =\frac{1}{i\pi^2}\int \frac{d^4 k}{D_0D_1D_2},
 \label{scalrInte}\eea
 where $n=1,2$, $D=4-2\epsilon \leq 4$ is the dimension of the integral, while $~M_0,~M_1,~M_2$ stand for the masses of virtual particles in the loops. We also assume  $p^2_1=m^2_{1},~p^2_2=m^2_{2}$ for external fermions. The tensor integrals are
 \bea
 A^{\mu}(p_n;M_n)
 &=&\frac{\left(2\pi\mu\right)^{4-D}}{i\pi^2}\int \frac{d^D k\times k^{\mu}}{D_n}=A_0(M_n)p_n^{\mu},\crn
 B^{\mu}(p_n;M_0,M_n)&=& \frac{\left(2\pi\mu\right)^{4-D}}{i\pi^2}\int \frac{d^D k\times
 	k^{\mu}}{D_0D_n}\equiv B^{(n)}_1p^{\mu}_n,\crn
 B^{\mu}(p_1,p_2;M_1,M_2)&=& \frac{\left(2\pi\mu\right)^{4-D}}{i\pi^2}\int \frac{d^D k\times
 	k^{\mu}}{D_1D_2}\equiv B^{(12)}_1p^{\mu}_1+B^{(12)}_2p^{\mu}_2,\crn
 C^{\mu}(M_0,M_1,M_2)&=&\frac{1}{i\pi^2}\int \frac{d^4 k\times k^{\mu}}{D_0D_1D_2}\equiv  C_1 p_1^{\mu}+C_2 p_2^{\mu},\crn
 C^{\mu \nu}(M_0,M_1,M_2)&=&\frac{1}{i\pi^2}\int \frac{d^4 k\times k^{\mu}k^{\nu}}{D_0D_1D_2}\equiv  C_{00}g^{\mu \nu}+C_{11} p_1^{\mu}p_1^{\nu}+C_{12} p_1^{\mu}p_2^{\nu}+C_{21} p_2^{\mu}p_1^{\nu}+C_{22} p_2^{\mu}p_2^{\nu},\crn
 \label{oneloopin1}\eea
 where $A_0$, $B^{(n)}_{0,1}$, $B^{(12)}_{n}$ and $C_{0,n}, C_{mn}$   are PV functions.  It is well-known that $C_{0,n}, C_{mn}$ are finite while the remains are divergent. We denote
 \be \Delta_{\epsilon}\equiv \frac{1}{\epsilon}+\ln4\pi-\gamma_E, \label{divt}\ee with $\gamma_E$ is the  Euler constant.  
 
 Using the technique as mentioned in Ref.\cite{Hue:2017lak}, we can show the divergent parts of the above PV functions as
 \bea  \mathrm{Div}[A_0(M_n)]&=& M_n^2 \Delta_{\epsilon}, \hs  \mathrm{Div}[B^{(n)}_0]= \mathrm{Div}[B^{(12)}_0]= \Delta_{\epsilon}, \crn
 \mathrm{Div}[B^{(1)}_1]&=&\mathrm{Div}[B^{(12)}_1] = \frac{1}{2}\Delta_{\epsilon},  \hs  \mathrm{Div}[B^{(2)}_1] = \mathrm{Div}[B^{(12)}_2]= -\frac{1}{2} \Delta_{\epsilon}.  \label{divs1}\eea
 Apart from the divergent parts, the rest of these functions are finite.
 
 Thus, the above PV functions can be written in form:
 \be  A_0(M)= M^2\Delta_{\epsilon}+a_0(M),\,\, B^{(n)}_{0,1}= \mathrm{Div}[B^{(n)}_{0,1}]+ b^{(n)}_{0,1}, \,\,  B^{(12)}_{0,1,2}= \mathrm{Div}[B^{(12)}_{0,1,2}]+ b^{(12)}_{0,1,2}, \label{B01i}\ee
 where $a_0(M), \,\,b^{(n)}_{0,1}, \,\, b^{(12)}_{0,1,2} $ are finite parts and have a specific form defined as Ref.\cite{Thuc:2016qva} for $\mathrm{H} \rightarrow \mu \tau$ decay.
 \section{Analytic formulas at one-loop order  to $\mathrm{H} \rightarrow l_il_j$.}
 \label{appen_loops1}
 For the convenience of analytic representation, we use the following notations: $\mathcal{A}_{L,R}^{(k)G}\equiv\mathcal{A}_{(ab)L,R}^{(k)G}$ ,\, $\mathcal{A}_{L,R}^{(k)H_s^\pm}\equiv\mathcal{A}_{(ab)L,R}^{(k)H_s^\pm}$,\,\, $G \equiv W,V-boson$,\, $H^\pm_s \equiv H^\pm_{1,2}$, \,$\mathrm{H} \equiv h^0_{1,2,3,4}$.
 \bea
 U^{\nu *}_G= \left\lbrace \begin{array}{cc}
 	U^{\nu *}_{ai} & \mathrm{if} \, G \equiv W \\
 	U^{\nu *}_{(a+3)i} & \mathrm{if} \, G \equiv V \\
 \end{array} \right.  \, \mathrm{and} \,\,  U^{\nu}_G= \left\lbrace \begin{array}{cc}
 U^{\nu}_{bi} & \mathrm{if} \, G \equiv W \\
 U^{\nu}_{(b+3)i} & \mathrm{if} \, G \equiv V \\
\end{array} \right. \nonumber
 \eea
 The Passarino-Veltman functions are given as shown in \cite{Passarino:1978jh}, and have a common set of variables ($p_k^2,m_1^2,m_2^2,m_3^2$) with $p_k$ are external momenta and $m_1^2,m_2^2,m_3^2$ related to masses in loop of figures \ref{fig_hmt}. We also use the formulas for determining the symmetry factors of Feynman diagrams as shown in Refs.\cite{Dong:2009db, Hue:2010xr}.
 
 The analytic expressions for $\mathcal{A}_{L,R}^{(k)G}$ and  $\mathcal{A}_{L,R}^{(k)H_s^\pm}$  where $k$ implies the diagram (k) in Fig. \ref{fig_hmt}, are
 \bea
 \mathcal{A}^{(1)G}_{L}\left(m_G,m_{\mathrm{H}} \right) & =& \frac{g^3m_{a}}{64\pi^2 m_G^3}\sum_{i=1}^{9}U^{\nu*}_GU^{\nu}_G
 \left\{ m_{n_i}^2\left(B^{(1)}_1- B^{(1)}_0- B^{(2)}_0\right) -m_b^2 B^{(2)}_1  +\left(2m_G^2+m^2_{\mathrm{H}}\right)m_{n_i}^2 C_0 \right.\crn &-&\left. \left[2m_G^2\left(2m_G^2+m_{n_i}^2+m_a^2-m_b^2\right) + m_{n_i}^2m_{\mathrm{H}}^2\right] C_1 +
 \left[2m_G^2\left(m_a^2-m^2_{\mathrm{H}}\right)+ m_b^2 m^2_{\mathrm{H}}\right]C_2\frac{}{}\right\},\crn
 \mathcal{A}^{(1)G}_{R}\left(m_G,m_{\mathrm{H}} \right)&= &\frac{g^3m_b}{64\pi^2 m_G^3}\sum_{i=1}^{9}U^{\nu*}_GU^{\nu}_G
 \left\{ -m_{n_i}^2\left(B^{(2)}_1+B^{(1)}_0+ B^{(2)}_0\right) +m_a^2 B^{(1)}_1  +\left(2m_G^2+m^2_{\mathrm{H}}\right)m_{n_i}^2 C_0 \right.\crn &-&\left.
 \left[2m_G^2\left(m_b^2-m^2_{h}\right)+ m_a^2 m^2_{\mathrm{H}}\right]C_1 + \left[2m_G^2\left(2m_G^2+m_{n_i}^2-m_a^2+m_b^2\right) + m_{n_i}^2m_{\mathrm{H}}^2\right] C_2 \frac{}{}\right\},\crn	
 \label{de1G}\eea
  \bea 
 \mathcal{A}^{(2)G}_{L}  \left(m_G,m_{\mathrm{H}},m_{H^\pm_s} \right)&=&\frac{g^3 m_{a}}{64 \pi^2 m_G^3}\sum_{i=1}^{9}U^{\nu*}_{G}\crn
 &\times&\left\{\lambda^{L,s}_{bi}m_{n_i}\left[ B^{(1)}_0-B^{(1)}_1+\left(m_G^2+m_{H^\pm_s}^2-m_{\mathrm{H}}^2\right)C_0+\left(m_G^2-m_{H^\pm_s}^2+m_{\mathrm{H}}^2\right)C_1\right]\right.\crn
 && +\left. \lambda^{R,s}_{bi}m_{b}\left[ 2m_G^2C_1-\left(m_G^2+m_{H^\pm_s}^2-m_{\mathrm{H}}^2\right)C_2\right]\right\},\crn
 \mathcal{A}^{(2)G}_{R}\left(m_G,m_{\mathrm{H}},m_{H^\pm_s} \right) &=&\frac{g^3}{64 \pi^2 m_G^3}\sum_{i=1}^{9}U^{\nu*}_{G}\crn
 &\times&\left\{\lambda^{L,s}_{bi}m_bm_{n_i}\left[ -2m_G^2C_0-\left(m_G^2-m_{H^\pm_s}^2+m_{\mathrm{H}}^2\right)C_2\right]\right.\crn
 &&+
 \left. \lambda^{R,s}_{bi}\left[-m_{n_i}^2 B^{(1)}_0+m_a^2B^{(1)}_1+m_{n_i}^2\left(m_G^2-m_{H^\pm_s}^2+m_{\mathrm{H}}^2\right)C_0\right.\right.\crn
 &&+\left.\left.\left[ 2m_G^2\left(m_{\mathrm{H}}^2-m_b^2\right)- m_a^2\left(m_G^2-m_{H^\pm_s}^2+m_{\mathrm{H}}^2\right)\right]C_1 +2 m_b^2m_G^2C_2\right]\right\}, \label{d2LR}
 \eea
  \bea
 \mathcal{A}^{(3)G}_{L}\left(m_G,m_{\mathrm{H}},m_{H^\pm_s} \right) &=&\frac{g^3 }{64\pi^2 m_G^3}\sum_{i=1}^{9}U^{\nu}_{G}\crn
 &\times&\left\{\lambda^{L,s*}_{ai}m_am_{n_i}\left[ -2m_G^2C_0+\left(m_G^2-m_{H^\pm_s}^2+m_{h^0_1}^2\right)C_1\right]\right.\crn
 &&+
 \left. \lambda^{R,s*}_{ai}\left[-m_{n_i}^2 B^{(2)}_0-m_b^2B^{(2)}_1+m_{n_i}^2\left(m_G^2-m_{H^\pm_s}^2+m_{\mathrm{H}}^2\right)C_0\right.\right.\crn
 &&-\left.\left. 2m_a^2m_Y^2C_1-\left[ 2m_G^2\left(m_{\mathrm{H}}^2-m_a^2\right)- m_b^2\left(m_G^2-m_{H^\pm_s}^2+m_{\mathrm{H}}^2\right)\right]C_2\right]\right\},\crn
 \mathcal{A}^{(3)G}_{R}\left(m_G,m_{\mathrm{H}},m_{H^\pm_s} \right) &=&\frac{g^3 m_b }{64\pi^2 m_G^3}\sum_{i=1}^{9}U^{\nu}_{G}\crn
 &\times&\left\{\lambda^{L,s*}_{ai}m_{n_i}\left[ B^{(2)}_0+B^{(2)}_1+\left(m_G^2+m_{H^\pm_s}^2-m_{\mathrm{H}}^2\right)C_0-\left(m_G^2-m_{H^\pm_s}^2+m_{\mathrm{H}}^2\right)C_2\right]\right.\crn
 && +\left. \lambda^{R,s*}_{ai}m_{a}\left[  \left(m_G^2+m_{H^\pm_s}^2-m_{\mathrm{H}}^2\right)C_1-2m_G^2C_2\right]\right\},
 \label{d3LR}
 \eea
 \bea
 \mathcal{A}^{(4+5)G}_{L}\left(m_G\right) &=& \frac{g^3m_am_b^2}{64\pi^2m^3_G(m_a^2-m_b^2)} \sum_{i=1}^{9}U^{\nu*}_{G}U^{\nu}_{G}\left[  2m_{n_i}^2\left(B^{(1)}_0-B^{(2)}_0\right) \right. \crn&-&\left. \left(2 m_G^2 +m_{n_i}^2\right) \left(B^{(1)}_1 +B^{(2)}_1 \right)- m_a^2 B^{(1)}_1 -m_b^2 B^{(2)}_1 \right],  
 \crn
 \mathcal{A}^{(4+5)G}_{R}\left(m_G\right) &=&\frac{m_a}{m_b}\mathcal{A}^{(4+5)G}_{L}\left(m_G\right), \label{de45LR}
 \eea
 \bea
 \mathcal{A}^{(6)H_s^\pm}_{L}\left(m_{H^\pm_s},m_{l_a} \right) &=&   -\frac{g^3}{64\pi^2  m_W^3}\sum_{i,j=1}^{9}\left\{\lambda^{\mathrm{H}*}_{ij}\left[\lambda^{R,s*}_{ai}\lambda^{L,s}_{bj}\left(B^{(12)}_0+m_{H^\pm_s}^2C_0 -m_a^2 C_1+m_b^2C_2\right)\right.\right.\crn
 &+&\left.\left. \lambda^{R,s*}_{ai}\lambda^{R,s}_{bj}m_bm_{n_j}C_2 -\lambda^{L,s*}_{ai}\lambda^{L,s}_{bj}m_am_{n_i}C_1 \right]\right. \crn
 &+&\left.  \lambda^{\mathrm{H}}_{ij}\left[\lambda^{R,s*}_{ai}\lambda^{L,s}_{bj}m_{n_i}m_{n_j}C_0 +\lambda^{R,s*}_{ai}\lambda^{R,s}_{bj}m_{n_i}m_{b}(C_0+C_2)\right.\right.\crn
 &+&\left.\left.\lambda^{L,s*}_{ai}\lambda^{L,s}_{bj}m_{a}m_{n_j}(C_0-C_1)+ \lambda^{L,s*}_{ai}\lambda^{R,s}_{bj}m_{a}m_{b}(C_0-C_1+C_2) \right]\frac{}{}\right\},\crn
 \mathcal{A}^{(6)H_s^\pm}_{R}\left(m_{H^\pm_s},m_{l_a} \right) &=&-\frac{g^3}{64\pi^2  m_W^3} \sum_{i,j=1}^{9}\left\{\lambda^{\mathrm{H}}_{ij}\left[\lambda^{L,s*}_{ai}\lambda^{R,s}_{bj}\left(B^{(12)}_0+m_{H^\pm_s}^2C_0 -m_a^2 C_1+m_b^2C_2\right)\right.\right.\crn
 &+&\left.\left.\lambda^{L,s*}_{ai}\lambda^{L,s}_{bj}m_bm_{n_j}C_2-  \lambda^{R,s*}_{ai}\lambda^{R,s}_{bj}m_am_{n_i}C_1 \right]\right.\crn
 &+&\left.
 \lambda^{\mathrm{H}*}_{ij}\left[\lambda^{L,s*}_{ai}\lambda^{R,s}_{bj}m_{n_i}m_{n_j}C_0 +\lambda^{L,s*}_{ai}\lambda^{L,s}_{bj}m_{n_i}m_{b}(C_0+C_2)\right.\right.\crn
 &+&\left.\left.\lambda^{R,s*}_{ai}\lambda^{R,s}_{bj}m_{a}m_{n_j}(C_0-C_1)+ \lambda^{R,s*}_{ai}\lambda^{L,s}_{bj}m_{a}m_{b}(C_0-C_1+C_2) \right]\frac{}{}\right\},\label{d6LR}
 \eea
 \bea
 \mathcal{A}^{(7)H^\pm_{s}}_{L}\left(m_{H^\pm_s} \right) &=&\frac{g^2\lambda^{\pm}_{H_s}}{64\pi^2  m_W^2}\sum_{i=1}^{9}\left[-\lambda^{R,s*}_{ai}\lambda^{L,s}_{bi}m_{n_i}C_0 -\lambda^{L,s*}_{ai}\lambda^{L,s}_{bi}m_{a}C_1 +\lambda^{R,s*}_{ai}\lambda^{R,s}_{bi}m_{b}C_2 \right],\crn
 \mathcal{A}^{(7)H^\pm_{s}}_{R}\left(m_{H^\pm_s} \right) &=&\frac{g^2\lambda^{\pm}_{H_s}}{64\pi^2 m_W^2}\sum_{i=1}^{9}\left[-\lambda^{L,s*}_{ai}\lambda^{R,s}_{bi}m_{n_i}C_0 -\lambda^{R,s*}_{ai}\lambda^{R,s}_{bi}m_{a}C_1 +\lambda^{L,s*}_{ai}\lambda^{L,s}_{bi}m_{b}C_2 \right], \label{d7LR}
 \eea
 \bea
 \mathcal{A}^{(8)G}_{L}\left(m_G,m_{l_a} \right)  &=& \frac{g^3m_a}{64\pi^2   m_G^3}\sum_{i,j=1}^{9}U^{\nu*}_{G}U^{\nu}_{G} \left\{\lambda^{\mathrm{H}*}_{ij}m_{n_j}\left[B^{(12)}_0-m_G^2C_0+\left(2 m_G^2+m_{n_i}^2-m_a^2\right)C_1\right]\right. \crn
 &&\hspace{4cm}\left.+\lambda^{\mathrm{H}}_{ij}m_{n_i}\left[B^{(1)}_1+\left(2 m_G^2+m_{n_j}^2-m_b^2\right)C_1\right]\right\},\crn
 \mathcal{A}^{(8)G}_{R}\left(m_G,m_{l_a} \right) &=&\frac{g^3 m_b}{64\pi^2 m_G^3}\sum_{i=1}^{9}U^{\nu*}_{G}U^{\nu}_{G} \left\{\lambda^{\mathrm{H}}_{ij}m_{n_i}\left[B^{(12)}_0-m_G^2C_0-\left(2 m_G^2+m_{n_j}^2-m_b^2\right)C_2\right]\right. \crn
 &&\hspace{3.7cm}-\left.\lambda^{\mathrm{H}*}_{ij}m_{n_j}\left[B^{(2)}_1+\left(2 m_G^2+m_{n_i}^2-m_a^2\right)C_2\right]\right\},\label{d8LR}
 \eea
 \bea
 \mathcal{A}^{(9+10)H_s^\pm}_{L} \left(m_{H^\pm_s,m_{l_a}} \right) &=&-\frac{g^3}{64\pi^2m_W^3\left(m_a^2-m_b^2\right)} \crn
 &\times&\sum_{i=1}^{9}\left[ m_am_bm_{n_i}  \lambda^{L,s*}_{ai}\lambda^{R,s}_{bi}\left(B^{(1)}_0-B^{(2)}_0\right)+ m_{n_i}
 \lambda^{R,s*}_{ai}\lambda^{L,s}_{bi}\left(m^2_bB^{(1)}_0-m^2_aB^{(2)}_0\right)\right.\crn
 &&\left.+ m_{a}m_b \left(\lambda^{L,s*}_{ai}\lambda^{L,s}_{bi}m_b + \lambda^{R,s*}_{ai}\lambda^{R,s}_{bi}m_a\right)\left(B^{(1)}_1+ B^{(2)}_1\right)\right],\crn
 \mathcal{A}^{(9+10)H_s^\pm}_{R} \left(m_{H^\pm_s,m_{l_a}} \right) &=&-\frac{g^3}{64\pi^2m_W^3\left(m_a^2-m_b^2\right)}\crn
 &\times&\sum_{i=1}^{9}\left[ m_am_bm_{n_i}  \lambda^{R,s*}_{ai}\lambda^{L,s}_{bi}\left(B^{(1)}_0-B^{(2)}_0\right)+ m_{n_i}
 \lambda^{L,s*}_{ai}\lambda^{R,s}_{bi}\left(m^2_bB^{(1)}_0-m^2_aB^{(2)}_0\right)\right.\crn
 &&\left.+ m_{a}m_b \left(\lambda^{R,s*}_{ai}\lambda^{R,s}_{bi}m_b + \lambda^{L,s*}_{ai}\lambda^{L,s}_{bi}m_a\right)\left(B^{(1)}_1+ B^{(2)}_1\right)\right].
 \label{d910LR} \eea 
 Obviously, with the 7th diagram of Fig.(\ref{fig_hmt}) we always get that $\mathcal{A}^{(7)}_{L,R}(a_1,a_2,v_1,v_2,m_F,m_H)$ are finite because $C_i$ ($i=0,1,2$) do not contain divergent functions.
 \section{\label{appen_loops2} The analytic formulas and divergent cancellation in amplitudes}
 Using the equalities $M^{\nu}=U^{\nu*}\hat{M^{\nu}}U^{\nu\dagger}$ and   Eq. (\ref{Lnu1}), we can prove that
 \bea &&\mathrm{div}\left[\mathcal{A}^{(1)W}_{L,R}\right] \sim \sum_{i=1}^9U^{\nu*}_{ai}U^{\nu}_{bi}m^2_{n_i}=\left[U^{\nu}(\hat{M}^{\nu})^2U^{\nu\dagger}\right]_{ba}=(m_D^*m_D^T)_{ba}
 =(m_D^\dagger m_D)_{ba},
 \crn
 &&\mathrm{div}\left[\mathcal{A}^{(1)V}_{L,R}\right] \sim \sum_{i=1}^9U^{\nu*}_{(a+3)i}U^{\nu}_{(b+3)i}m^2_{n_i}=(m_D^\dagger m_D+M_R^*M_R^T)_{ba}\nn
 \eea
 To have the general formula for the above two forms, we give notation:
 \bea
 \zeta_G= \left\lbrace \begin{array}{cc}
 	$1$ & \mathrm{if} \, G \, \equiv V \\
 	$0$ & \mathrm{if} \, G \, \equiv  W \\
 \end{array} \right. \nonumber
\eea
 Therefore,
 \bea
 \mathrm{div}\left[\mathcal{A}^{(1)G}_{L,R}\right] &\sim &\sum_{i=1}^9U^{\nu*}_{G}U^{\nu}_{G}m^2_{n_i}=(m_D^\dagger m_D+\zeta_G M_R^*M_R^T)_{ba}.
 \eea
 Similarly, we can give the divergence parts for the formulas in App. \ref{appen_loops1} as:
 \bea
 \mathrm{div}\left[\mathcal{A}^{(2)G}_{L,R}\right] &\sim &\sum_{i=1}^9U^{\nu*}_{G}\lambda_{bi}^{L,s}m_{n_i}=(-m_D^\dagger m_D+ t_{13}^2\zeta_GM_R^*M_R^T)_{ba},\crn
 \mathrm{div}\left[\mathcal{A}^{(3)G}_{L,R}\right] &\sim &\sum_{i=1}^9U^{\nu}_{G}\lambda_{ai}^{L,s*}m_{n_i}=(m_D^\dagger m_D+ t_{23}^2\zeta_GM_R^*M_R^T)_{ba},\crn
  \mathrm{div}\left[\mathcal{A}^{(4+5)G}_{L,R}\right] &=&0, \crn
 \mathrm{div}\left[\mathcal{A}^{(6)H^\pm_{s}}_{L,R}\right] &\sim &\sum_{i,j=1}^9U^{\nu*}_{G}\lambda^{\mathrm{H}*}_{ij}\lambda^{R,s*}_{ai}\lambda^{L,s}_{bj}\sim (m_D^\dagger m_D+f^{\mathrm{H}}f^{L,s}M_R^*M_R^T)_{ba},\crn
 \mathrm{div}\left[\mathcal{A}^{(7)H^\pm_{s}}_{L,R}\right] &=&0,\crn
 \mathrm{div}\left[\mathcal{A}^{(8)G}_{L,R}\right] &\sim &\sum_{i,j=1}^9U^{\nu*}_{G}U^{\nu}_{G}\left( \lambda^{\mathrm{H}*}_{ij}m_{n_i}+\frac{1}{2}\lambda^{\mathrm{H}}_{ij}m_{n_j}\right) \sim (m_D^\dagger m_D+f^{\mathrm{H}}\zeta_GM_R^*M_R^T)_{ba},\crn
 \mathrm{div}\left[\mathcal{A}^{(9+10)H^\pm_{s}}_{L,R}\right] &\sim &\sum_{i=1}^9\lambda^{R,s*}_{ai}\lambda^{L,s}_{bj}m_{n_i}\sim (m_D^\dagger m_D+f^{L,s}\zeta_GM_R^*M_R^T)_{ba}.
 \eea
 where $f^{L,s}$ are the coefficients given from the couplings ($\lambda^{L,s}_{ij}$) of the charged Higgs bosons with the leptons, while $f^{\mathrm{H}}$ depends on the coefficients ($\lambda_{ij}^H$).\\
 
 As the above result, we have given the decay amplitude for the case of $h^0_1 \rightarrow l_il_j$. In a similar way, we will give the decay amplitudes for the remaining cases of $h^0_2$, $h^0_3$ and $h^0_4$. We use the results of the analysis for each diagram as given in App.\ref{appen_loops1} to represent the total amplitude. The specific results are:\\
 For $h^0_2 \rightarrow l_il_j$,
 \bea \Delta_{(ab)L,R}^{h^0_2} &&= -s_\alpha \times \mathcal{A}^{(1)W}_{L,R}\left( m_W\right)  -\frac{\left(\sqrt{2}c_\alpha c_{23} +s_\alpha s_{23} \right)}{\sqrt{2}} \times \mathcal{A}^{(1)V}_{L,R}\left( m_V\right) \crn
 && +\frac{\left(\sqrt{2}c_\alpha s_{23} -s_\alpha c_{23} \right)}{\sqrt{2}t_{23}} \times \mathcal{A}^{(2+3)V}_{L,R}\left( m_V,m_{H^\pm_2}\right)  +s_{\alpha}\times \mathcal{A}^{(4+5)W}_{L,R}\left( m_{W}\right)  \crn
 &&+\frac{s_{\alpha}}{s_{23}}\times \mathcal{A}^{(4+5)V}_{L,R}\left( m_{V}\right)+s_{\alpha}\times\mathcal{A}^{(6)H^\pm_1}_{L,R}\left( m_{H^\pm_1}\right)+2s_{\alpha}c_{23}^2\times \mathcal{A}^{(6)H^\pm_2}_{L,R}\left( m_{H^\pm_2}\right)  \crn
 &&+ 2\times \mathcal{A}^{(7)H^\pm_1}_{L,R}\left( m_{H^\pm_1}\right) + 4c^2_{23}\times \mathcal{A}^{(7)H^\pm_2}_{L,R}\left( m_{H^\pm_2}\right) - s_{\alpha}\times\mathcal{A}^{(8)W}_{L,R}\left( m_{W}\right) \crn
 && -\frac{s_{\alpha}}{s_{23}}\times \mathcal{A}^{(8)V}_{L,R}\left( m_{V}\right)+ s_{\alpha}\times \mathcal{A}^{(9+10)H^\pm_1}_{L,R}\left( m_{H^\pm_1}\right) + 2s_{\alpha}c_{23}^2 \times \mathcal{A}^{(9+10)H^\pm_2}_{L,R}\left( m_{H^\pm_2}\right),  \label{deLR2}
 \eea
 and we can check the canceling divergence for the right-components of the amplitude $\Delta_{(ab)}^{h^0_2}$ as,
 \begin{align}
 	& \mathrm{div}\left[\Delta^{(1)W}_R\right]+\mathrm{div}\left[\Delta^{(8)W}_R\right]=(m_D^\dagger m_D)_{ba} \left( -\frac{3}{2}s_\alpha +\frac{3}{2}s_\alpha  \right) = 0\crn
 	&\mathrm{div}\left[\Delta^{(6)H^\pm_1}_R\right]+\mathrm{div}\left[\Delta^{(9+10)H^\pm_1}_R\right]=(m_D^\dagger m_D)_{ba} \left( s_\alpha -s_\alpha  \right)= 0 \crn
 	& \mathrm{div}\left[\Delta^{(1+2+3+8)V}_R+\Delta^{(6+9+10)H^\pm_2}_R \right]\crn
 	&\sim (m_D^\dagger m_D)_{ba} \left\{\sqrt{2}s_{\alpha}c^2_{23}s_{23}(3-1-2) + c_{\alpha}\left[c^2_{23}(-3c^2_{23}-s^2_{23}-2s^2_{23}+3) +2 s^2_{23}-2s^2_{23} \right]\right\}\crn
 	&+ (M^*_RM^T_R)_{ba}\left[ \sqrt{2}s_{\alpha}\frac{c^2_{23}}{s_{23}} \left(3s^2_{23}+c^2_{23}-2s^2_{23}-3+2\right)+ c_{\alpha}s^2_{23}\left(-3s^2_{23}+s^2_{23}-2c^2_{23}+2\right)\right]\crn
 	&=0. \label{sdiv2}
 \end{align}
 For $h^0_3 \rightarrow l_il_j$,
 \bea \Delta_{(ab)L,R}^{h^0_3} &&= -\frac{s_{23}}{\sqrt{2}} \times \mathcal{A}^{(1)V}_{L,R}\left( m_V\right)+\frac{1}{2s^2_{23}} \times \mathcal{A}^{(2+3)W}_{L,R}\left( m_W,m_{H^\pm_1}\right)  \crn
 && +\frac{c^2_{23}}{s_{23}} \times \mathcal{A}^{(2+3)V}_{L,R}\left( m_V,m_{H^\pm_2}\right)  +\mathcal{A}^{(4+5)W}_{L,R}\left( m_{W}\right)  \crn
 &&+\frac{1}{s_{23}}\times \mathcal{A}^{(4+5)V}_{L,R}\left( m_{V}\right)+\mathcal{A}^{(6)H^\pm_1}_{L,R}\left( m_{H^\pm_1}\right)+2c_{23}^2\times \mathcal{A}^{(6)H^\pm_2}_{L,R}\left( m_{H^\pm_2}\right)  \crn
 &&+ 4c^2_{23}\times \mathcal{A}^{(7)H^\pm_2}_{L,R}\left( m_{H^\pm_2}\right) - \mathcal{A}^{(8)W}_{L,R}\left( m_{W}\right) \crn
 && -\frac{1}{s_{23}}\times \mathcal{A}^{(8)V}_{L,R}\left( m_{V}\right)+ \mathcal{A}^{(9+10)H^\pm_1}_{L,R}\left( m_{H^\pm_1}\right) + 2c_{23}^2 \times \mathcal{A}^{(9+10)H^\pm_2}_{L,R}\left( m_{H^\pm_2}\right).  \label{deLR3}
 \eea
 The devergence of right-components of $\Delta_{(ab)}^{h^0_3}$ are canceled in the following way:
 \begin{align}
 	& \mathrm{div}\left[\Delta^{(2+3)W}_R\right]+\mathrm{div}\left[\Delta^{(8)W}_R\right]=(m_D^\dagger m_D)_{ba} \left( -\frac{1}{2} +\frac{1}{2}\right) = 0\crn
 	&\mathrm{div}\left[\Delta^{(6)H^\pm_1}_R\right]+\mathrm{div}\left[\Delta^{(9+10)H^\pm_1}_R\right]=(m_D^\dagger m_D)_{ba} \left( \frac{1}{2} -\frac{1}{2}  \right)= 0 \crn
 	& \mathrm{div}\left[\Delta^{(1+2+3+8)V}_R+\Delta^{(6+9+10)H^\pm_2}_R \right]\crn
 	&\sim (m_D^\dagger m_D)_{ba} \left\{\sqrt{2}c^2_{23}s_{23}(3-1-2) + \left[c^2_{23}(-3c^2_{23}-s^2_{23}-2s^2_{23}+3) +2 s^2_{23}-2s^2_{23} \right]\right\}\crn
 	&+ (M^*_RM^T_R)_{ba}\left[ \sqrt{2}\frac{c^2_{23}}{s_{23}} \left(3s^2_{23}+c^2_{23}-2s^2_{23}-3+2\right)+ s^2_{23}\left(-3s^2_{23}+s^2_{23}-2c^2_{23}+2\right)\right]=0. \label{sdiv3}
 \end{align}
 For $h^0_4 \rightarrow l_il_j$,
 \bea 
 \Delta_{(ab)L,R}^{h^0_4} &&= c_{23}s_{23} \times \mathcal{A}^{(2+3)V}_{L,R}\left( m_V,m_{H^\pm_1}\right)+ \sqrt{2}c_{23}s^2_{23} \times \mathcal{A}^{(2+3)W}_{L,R}\left(m_W,m_{H^\pm_2}\right),\crn
 &&+2\sqrt{2}c_{23}\times \mathcal{A}^{(7)H^\pm_{1,2}}_{L,R}\left(m_{H^\pm_1}, m_{H^\pm_2}\right),  \label{deLR4}\eea
 and the canceling divergence are performed as,
 \begin{align}
 	&\mathrm{div}\left[\Delta^{(2+3)W}_R\right]=(m_D^\dagger m_D)_{ba} \frac{c_{23}s_{23}}{2}\left( -\frac{1}{2} +\frac{1}{2}\right) = 0\crn
 	&\mathrm{div}\left[\Delta^{(2+3)V}_R\right]=\left( (m_D^\dagger m_D)_{ba}+(M^*_RM^T_R)_{ba}\right)\frac{c_{23}s^2_{23}}{2}\left( -\frac{1}{2} +\frac{1}{2}\right) = 0. \label{sdiv4}
 \end{align}
This technique have shown in Refs.{\cite{Nguyen:2018rlb,Hung:2021fzb}}, we can do the same for the left-components of the amplitudes. 
 \section{\label{appen_loops3} Plots related to partial width of CP-even Higgs bosons in some other cases of $M_R$ diagonal matrix.}
\begin{figure}[ht]
 	\centering
 	\begin{tabular}{cc}
 	\includegraphics[width=6.5cm]{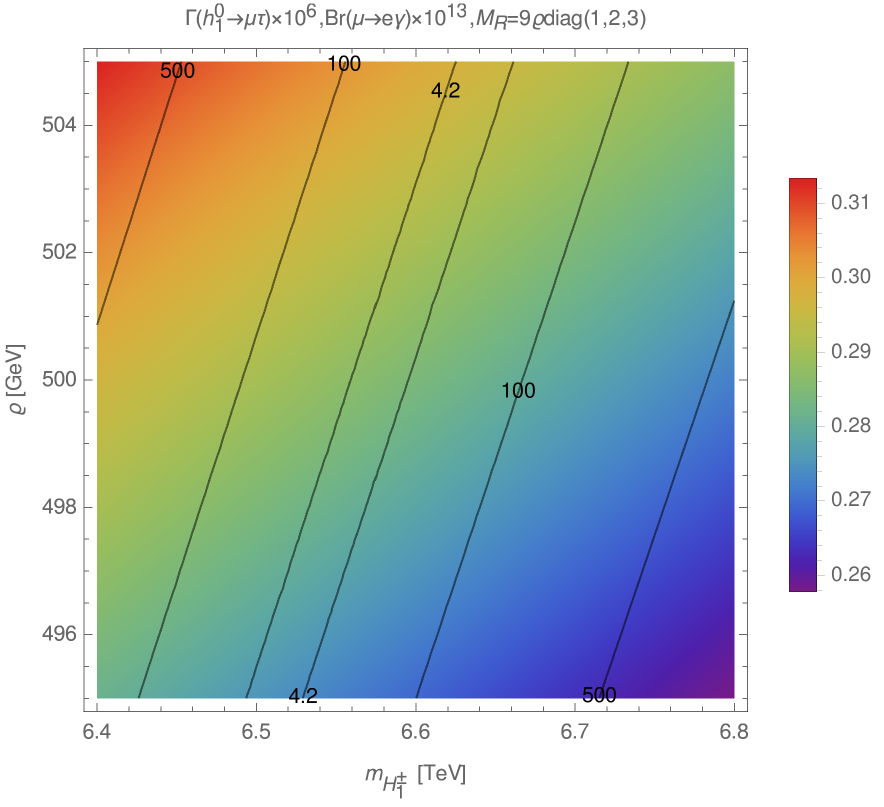} & \includegraphics[width=6.5cm]{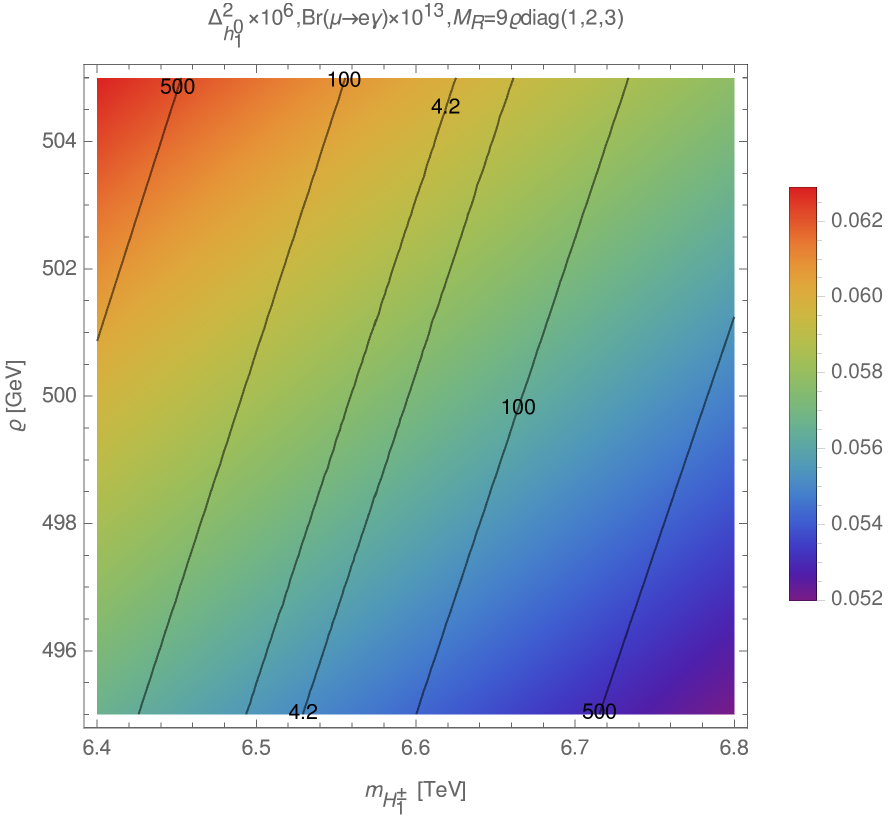}\\ 
 	\includegraphics[width=6.5cm]{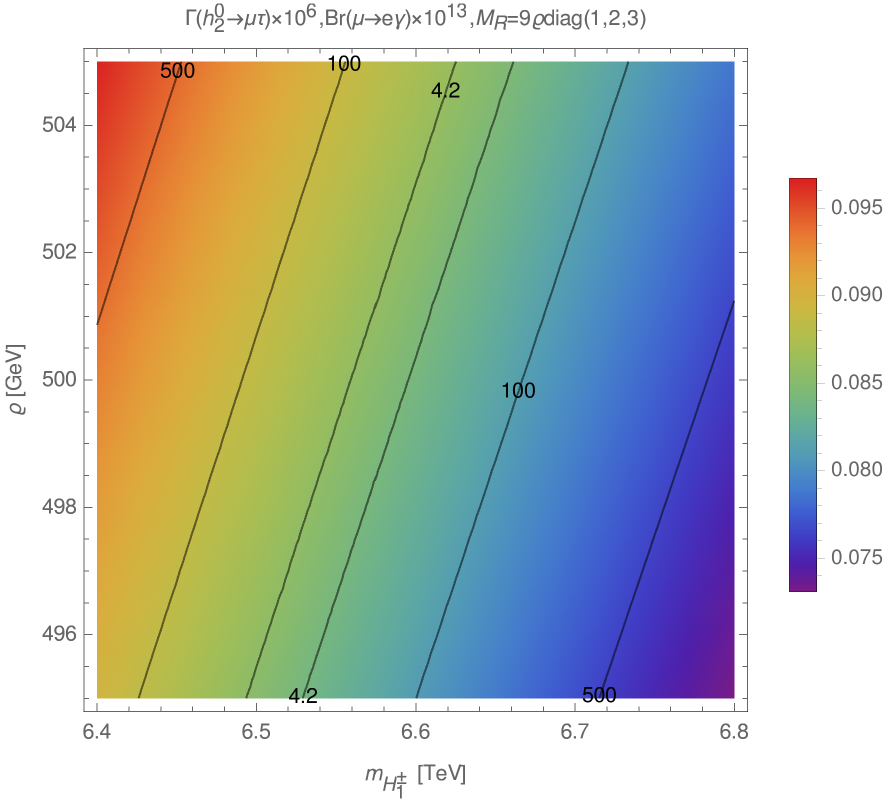} & \includegraphics[width=6.5cm]{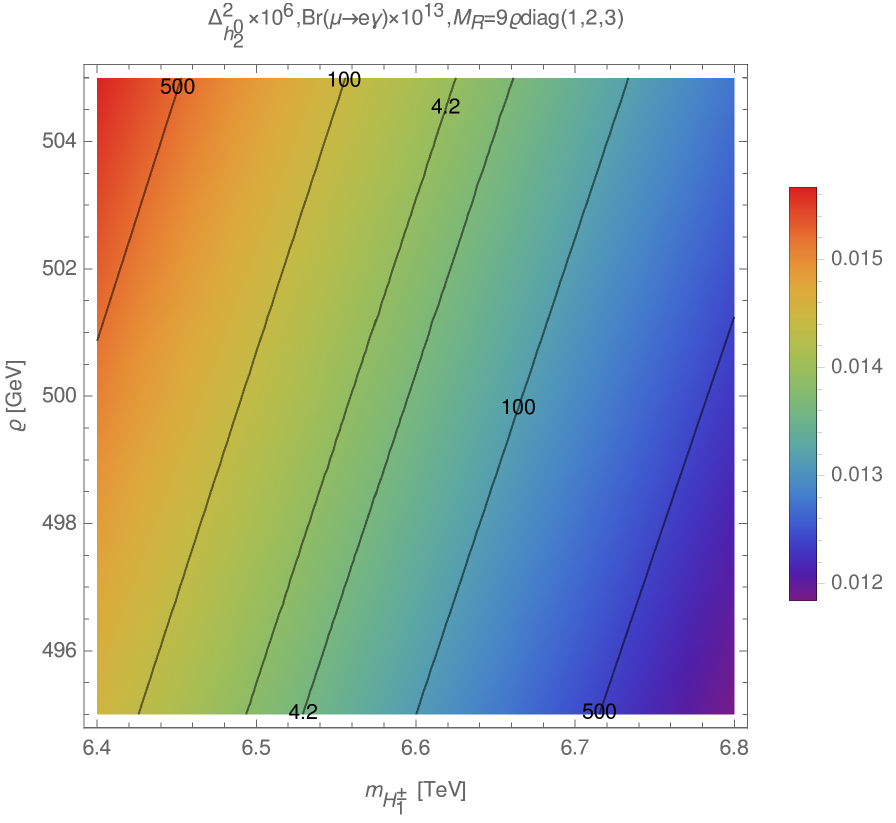}\\
 	\includegraphics[width=6.5cm]{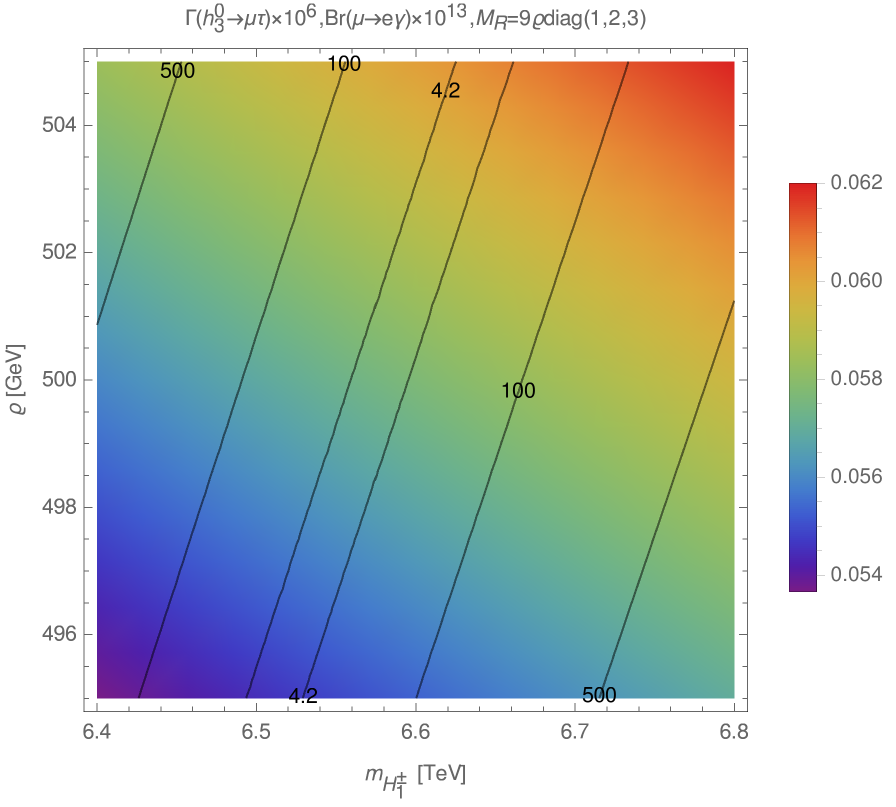} & \includegraphics[width=6.5cm]{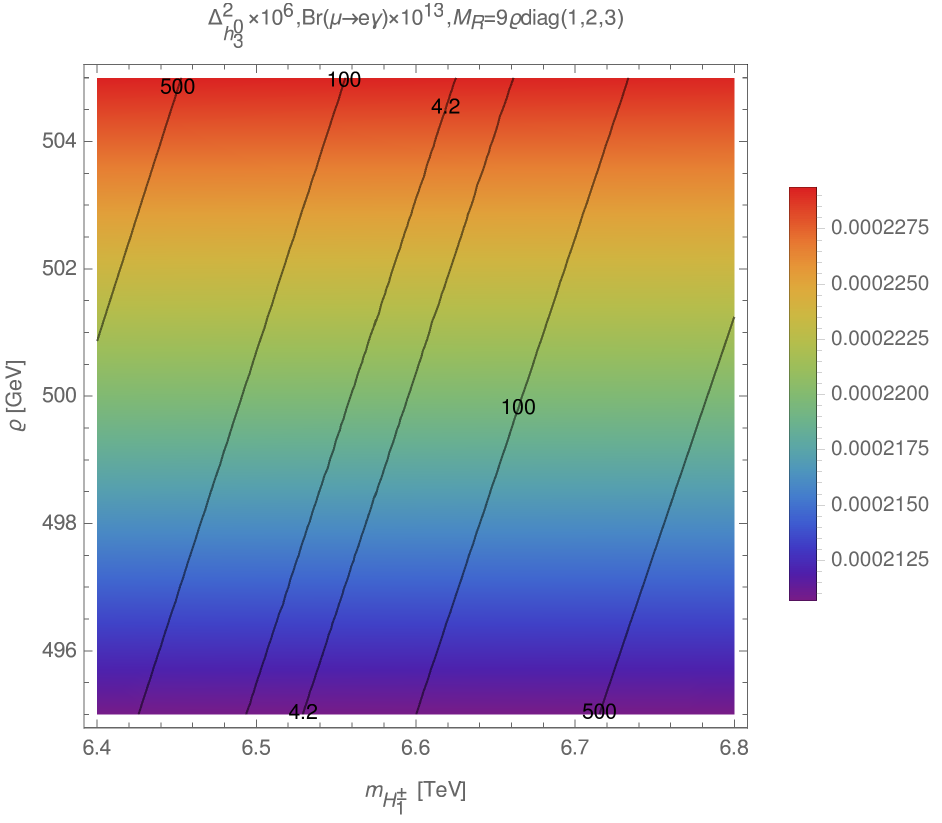}\\ 
 	\includegraphics[width=6.5cm]{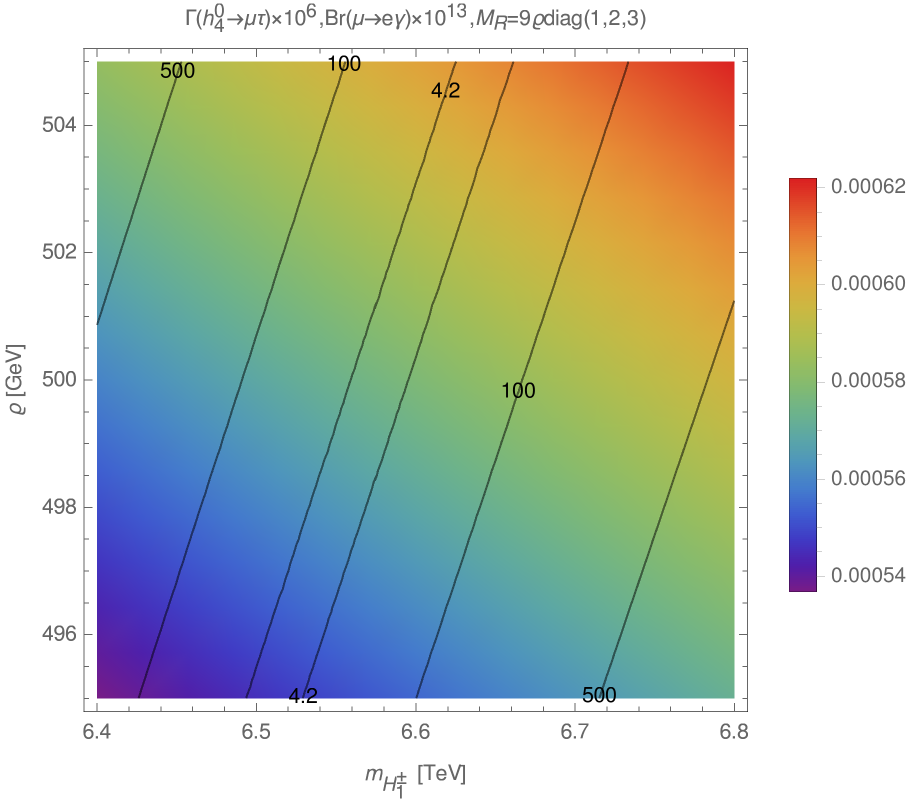} & \includegraphics[width=6.5cm]{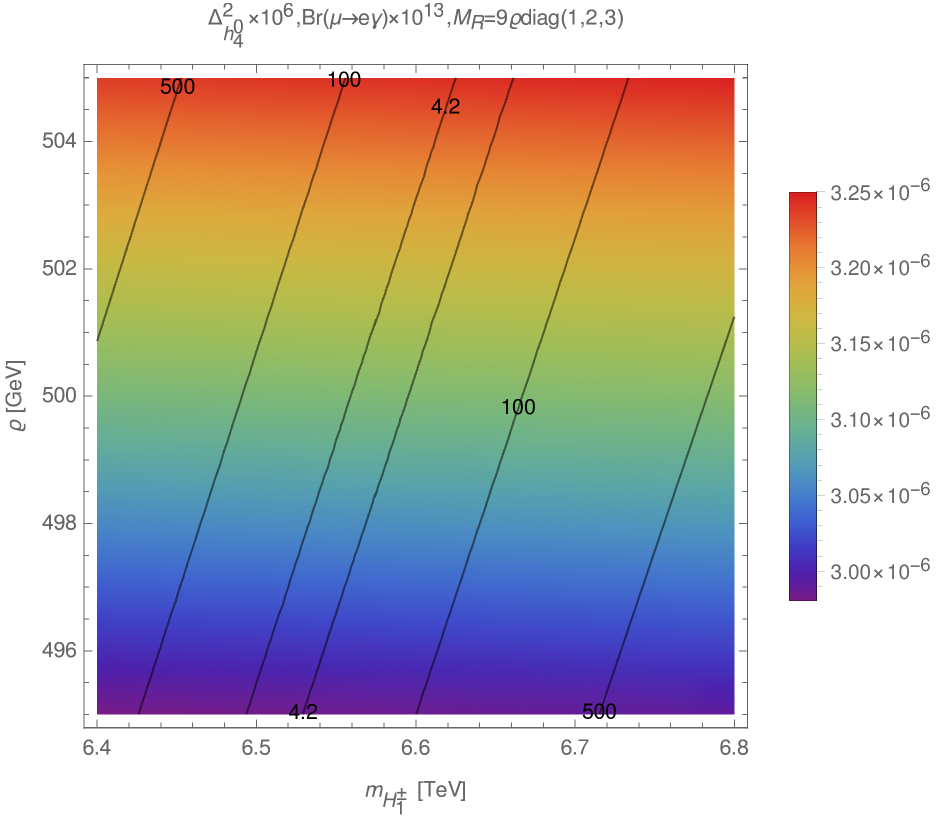}\\  
 	\end{tabular}%
 	\caption{ Density plots of $\Gamma (H\rightarrow \mu \tau)$ (left panel), $\Delta^2_{H} (H \rightarrow \mu \tau)$ (right panel) in the case of $M_R=9\varrho diag(1,2,3)$.}
 	\label{fig_Hmt123}
 \end{figure}

 Similar to the case of $M_R=9\varrho diag(1,1,1)$ presented above, we present numerical investigation results related to $\Gamma (H \rightarrow \mu \tau)$ and $\Delta^2_H (H \rightarrow \mu \tau)$ in this section.\\
 For the case of $M_R=9\varrho diag(1,2,3)$, we have plots in Fig.\ref{fig_Hmt123},\\
 For the case of $M_R=9\varrho diag(3,2,1)$, we obtain plots as Fig.\ref{fig_Hmt321},\\
 \small{\begin{figure}[ht]
 	\centering
 	\begin{tabular}{cc}
 	\includegraphics[width=6.5cm]{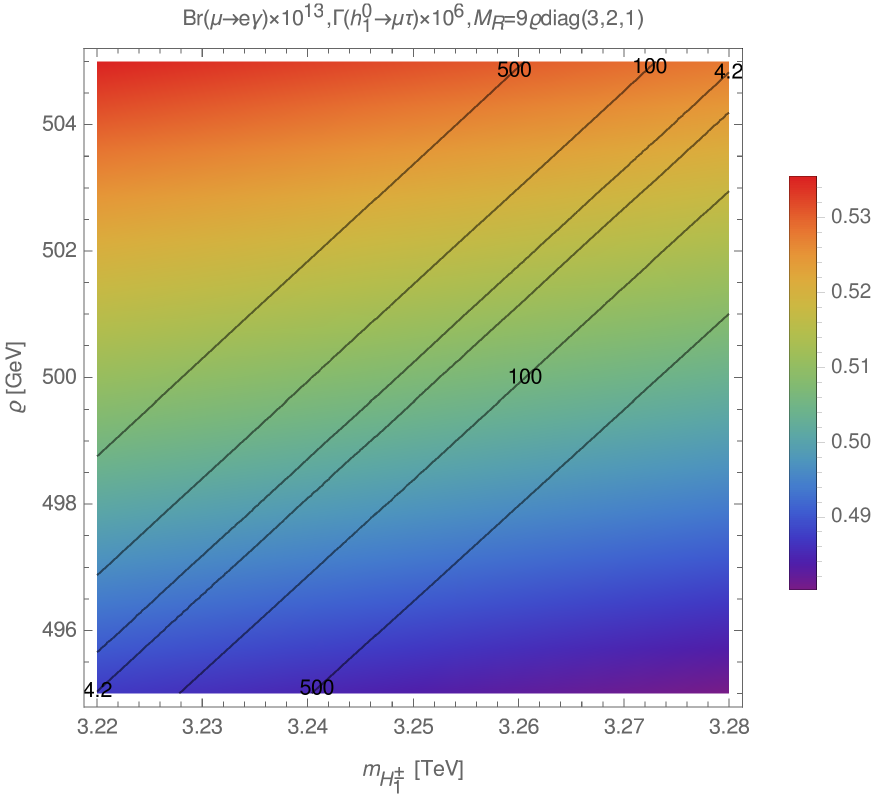} & \includegraphics[width=6.5cm]{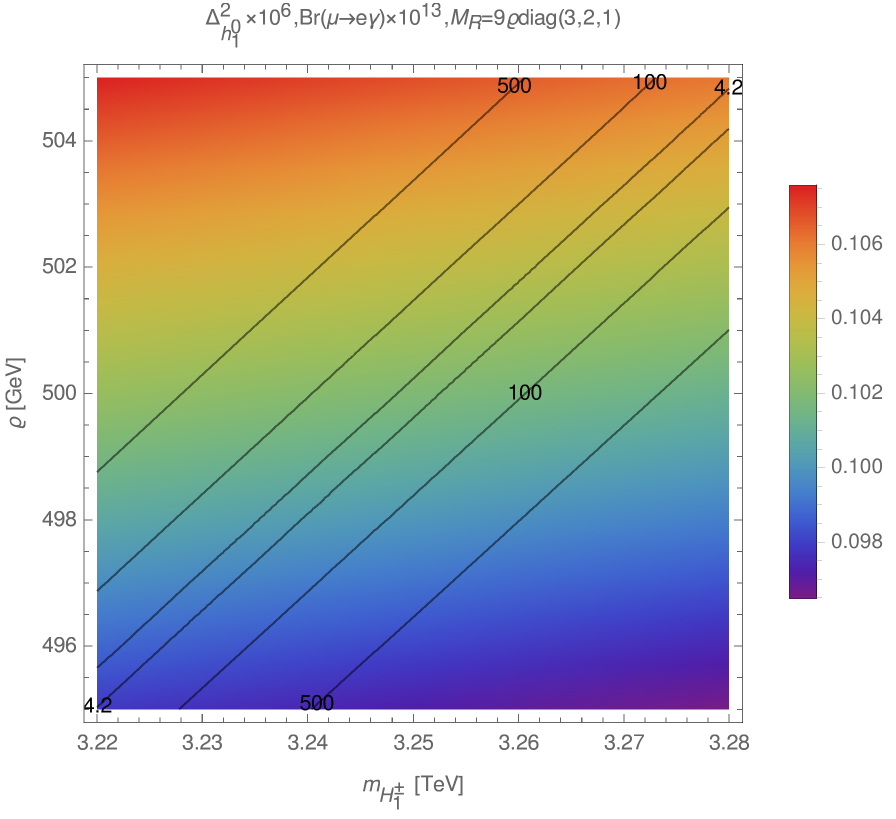}\\ 
 	\includegraphics[width=6.5cm]{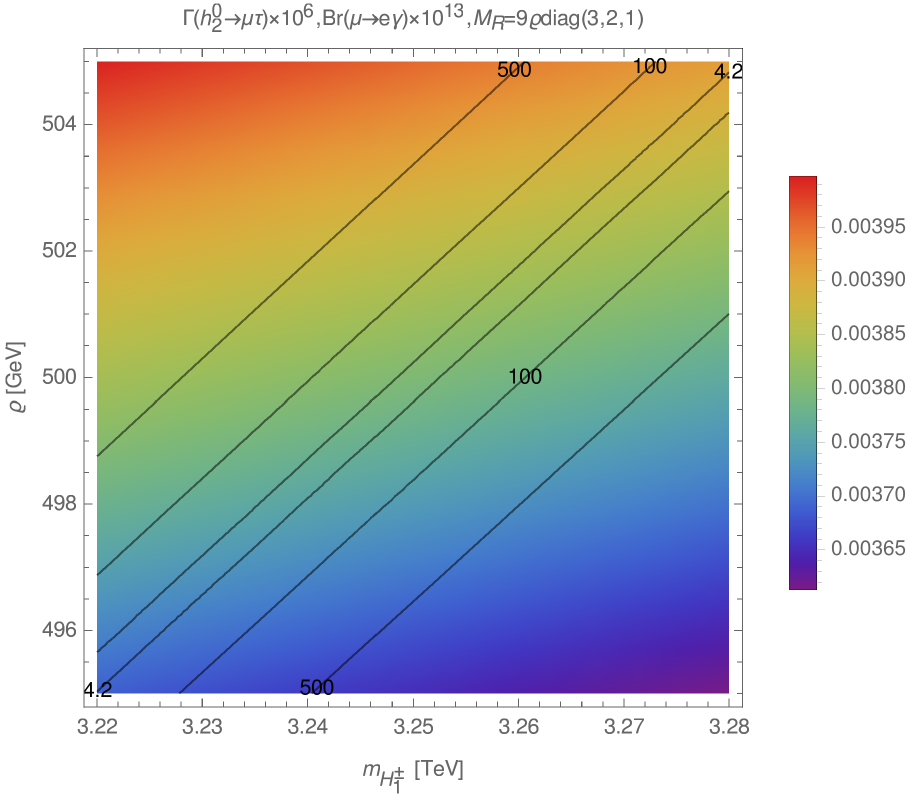} & \includegraphics[width=6.5cm]{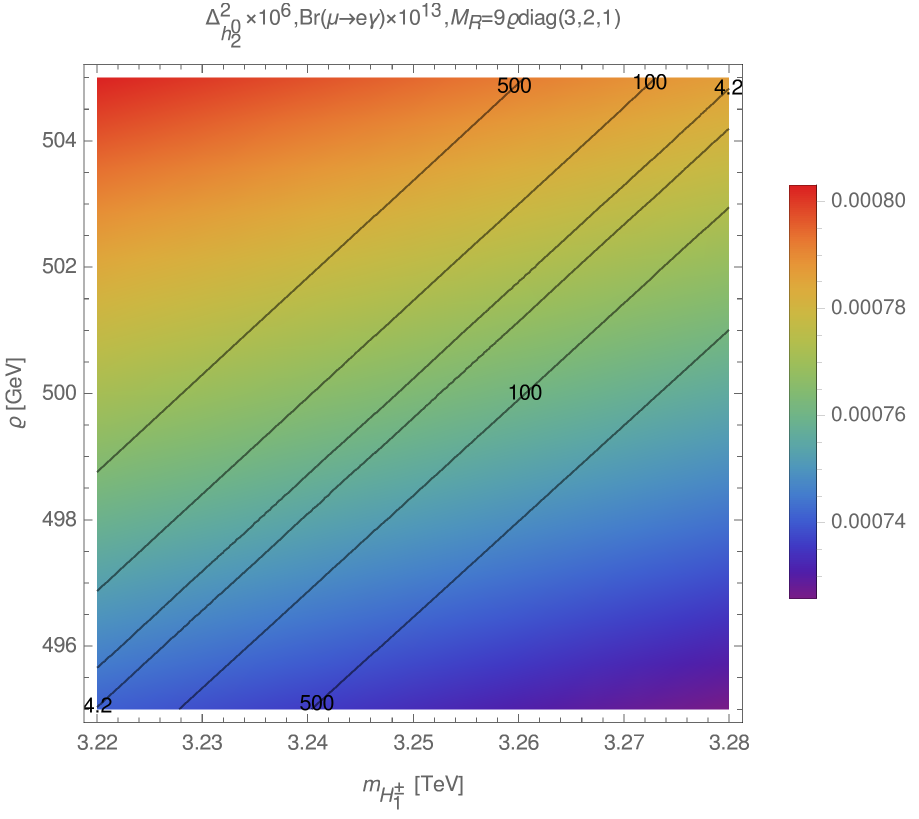}\\ 
 	\includegraphics[width=6.5cm]{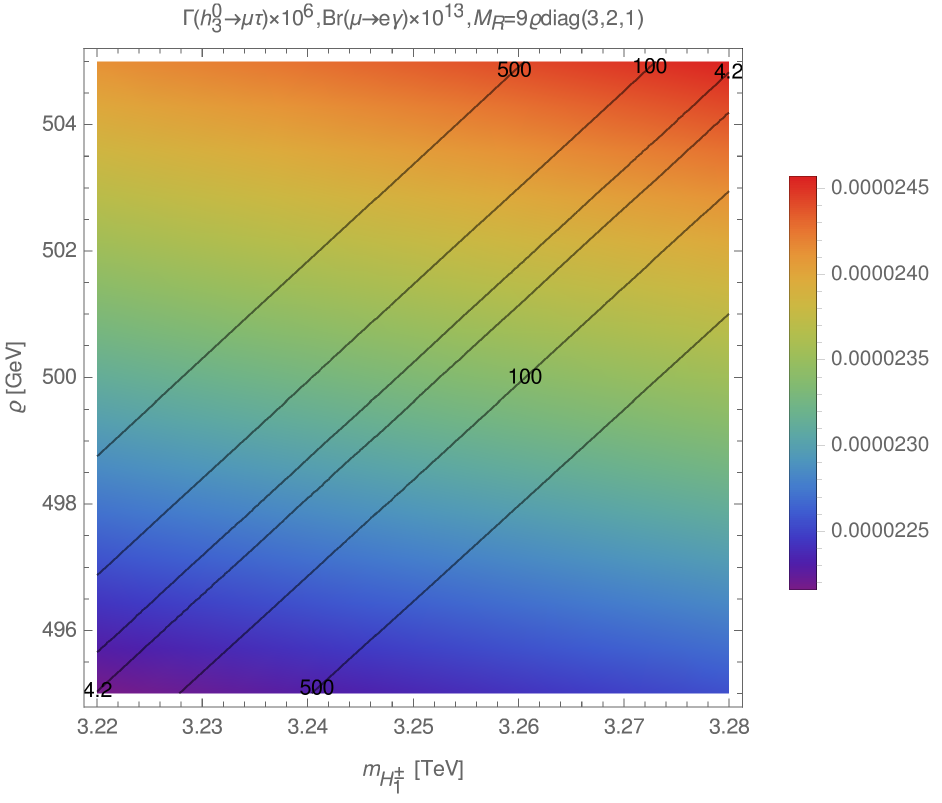} & \includegraphics[width=6.5cm]{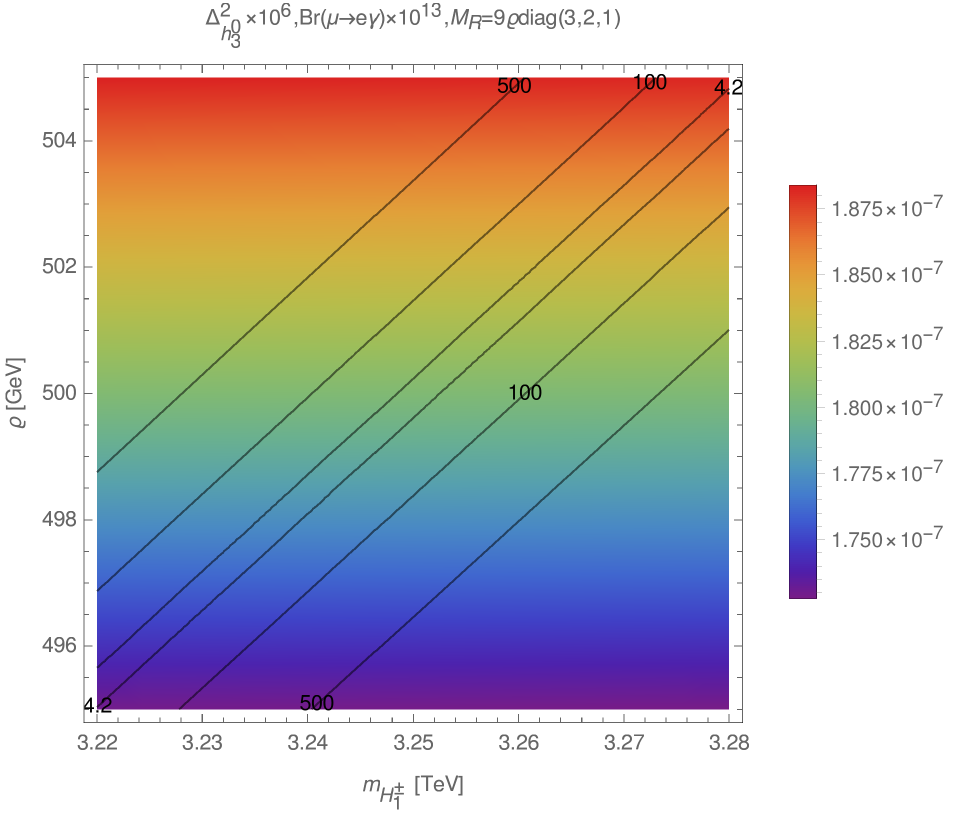}\\ 
 	\includegraphics[width=6.5cm]{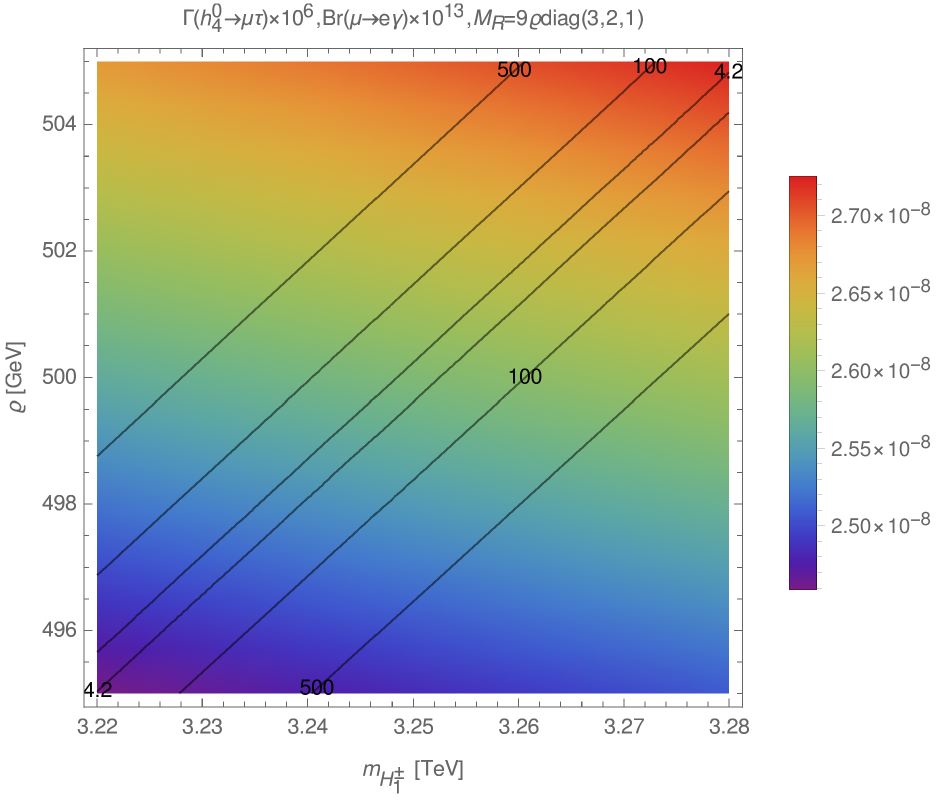} & \includegraphics[width=6.5cm]{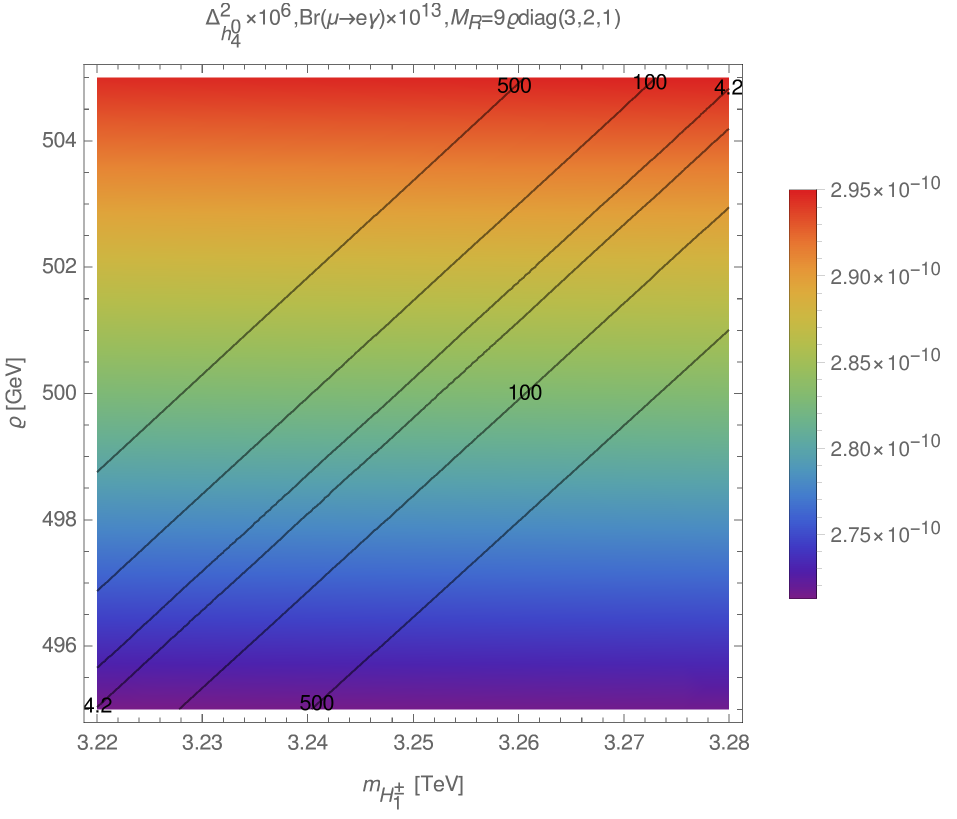}\\
 	\end{tabular}%
 	\caption{ Density plots of $\Gamma (H\rightarrow \mu \tau)$ (left panel), $\Delta^2_{H} (H \rightarrow \mu \tau)$ (right panel) in the case of $M_R=9\varrho diag(3,2,1)$.}
 	\label{fig_Hmt321}
 \end{figure}}

 The plots are all represented in the parameter space region satisfying the current experimental limit of $l_a \rightarrow l_b \gamma$ and all partial decay widths must satisfy $\Gamma (H \rightarrow \mu \tau)< 4.1 \times 10^{-6}$.

\end{document}